\documentstyle[twoside,epsfig]{article}

\catcode`\@=11
\long\def\@makefntext#1{
\protect\noindent \hbox to 3.2pt {\hskip-.9pt  
$^{{\eightrm\@thefnmark}}$\hfil}#1\hfill}		

\def\@makefnmark{\hbox to 0pt{$^{\@thefnmark}$\hss}}	
        
\def\ps@myheadings{\let\@mkboth\@gobbletwo
\def\@oddhead{\hbox{}
\rightmark\hfil\eightrm\thepage}   
\def\@oddfoot{}\def\@evenhead{\eightrm\thepage\hfil
\leftmark\hbox{}}\def\@evenfoot{}
\def\sectionmark##1{}\def\subsectionmark##1{}}

\oddsidemargin=\evensidemargin
\newcounter{sectionc}\newcounter{subsectionc}\newcounter{subsubsectionc}
\renewcommand{\section}[1] {\vspace{12pt}\addtocounter{sectionc}{1} 
\setcounter{subsectionc}{0}\setcounter{subsubsectionc}{0}\noindent 
        {\tenbf\thesectionc. #1}\par\vspace{5pt}}
\renewcommand{\subsection}[1] {\vspace{12pt}\addtocounter{subsectionc}{1} 
        \setcounter{subsubsectionc}{0}\noindent 
        {\bf\thesectionc.\thesubsectionc. {\kern1pt \bfit #1}}\par\vspace{5pt}}
\renewcommand{\subsubsection}[1] {\vspace{12pt}\addtocounter{subsubsectionc}{1}
        \noindent{\tenrm\thesectionc.\thesubsectionc.\thesubsubsectionc.
        {\kern1pt \tenit #1}}\par\vspace{5pt}}
\newcommand{\nonumsection}[1] {\vspace{12pt}\noindent{\tenbf #1}
        \par\vspace{5pt}}

\newcounter{appendixc}
\newcounter{subappendixc}[appendixc]
\newcounter{subsubappendixc}[subappendixc]
\renewcommand{\thesubappendixc}{\Alph{appendixc}.\arabic{subappendixc}}
\renewcommand{\thesubsubappendixc}
	{\Alph{appendixc}.\arabic{subappendixc}.\arabic{subsubappendixc}}

\renewcommand{\appendix}[1] {\vspace{12pt}
        \refstepcounter{appendixc}
        \setcounter{figure}{0}
        \setcounter{table}{0}
        \setcounter{lemma}{0}
        \setcounter{theorem}{0}
        \setcounter{corollary}{0}
        \setcounter{definition}{0}
        \setcounter{equation}{0}
        \renewcommand{\thefigure}{\Alph{appendixc}.\arabic{figure}}
        \renewcommand{\thetable}{\Alph{appendixc}.\arabic{table}}
        \renewcommand{\theappendixc}{\Alph{appendixc}}
        \renewcommand{\thelemma}{\Alph{appendixc}.\arabic{lemma}}
        \renewcommand{\thetheorem}{\Alph{appendixc}.\arabic{theorem}}
        \renewcommand{\thedefinition}{\Alph{appendixc}.\arabic{definition}}
        \renewcommand{\thecorollary}{\Alph{appendixc}.\arabic{corollary}}
        \renewcommand{\theequation}{\Alph{appendixc}.\arabic{equation}}
        \noindent{\tenbf Appendix \theappendixc #1}\par\vspace{5pt}}
\newcommand{\subappendix}[1] {\vspace{12pt}
        \refstepcounter{subappendixc}
        \noindent{\bf Appendix \thesubappendixc. {\kern1pt \bfit #1}}
	\par\vspace{5pt}}
\newcommand{\subsubappendix}[1] {\vspace{12pt}
        \refstepcounter{subsubappendixc}
        \noindent{\rm Appendix \thesubsubappendixc. {\kern1pt \tenit #1}}
	\par\vspace{5pt}}

\newcommand{\textlineskip}{\baselineskip=13pt}
\newcommand{\smalllineskip}{\baselineskip=10pt}

\def\eightcirc{
\begin{picture}(0,0)
\put(4.4,1.8){\circle{6.5}}
\end{picture}}
\def\eightcopyright{\eightcirc\kern2.7pt\hbox{\eightrm c}}

\newcommand{\pub}[1]{{\begin{center}\footnotesize\smalllineskip 
        Received #1\\
        \end{center}
        }}

\def\abstracts#1#2#3{{
        \centering{\begin{minipage}{4.5in}\baselineskip=10pt\footnotesize
        \parindent=0pt #1\par
        \parindent=15pt #2\par
        \parindent=15pt #3\par
        \end{minipage}}\par}} 

\def\keywords#1{{
       \centering{\begin{minipage}{4.5in}\baselineskip=10pt\footnotesize
       {\footnotesize\it Keywords}\/: #1
        \end{minipage}}\par}}

\renewenvironment{thebibliography}[1]
        {\frenchspacing
	 \ninerm\baselineskip=11pt
         \begin{list}{\arabic{enumi}.}
        {\usecounter{enumi}\setlength{\parsep}{0pt}     
         \setlength{\leftmargin 17pt}{\rightmargin 0pt}   
         \setlength{\itemsep}{0pt} \settowidth
	{\labelwidth}{#1.}\sloppy}}{\end{list}}

\newcounter{itemlistc}
\newcounter{romanlistc}
\newcounter{alphlistc}
\newcounter{arabiclistc}

\newcommand{\fcaption}[1]{
        \refstepcounter{figure}
	\setbox\@tempboxa = \hbox{\footnotesize Fig.~\thefigure. #1}
	\ifdim \wd\@tempboxa > 5in
           {\begin{center}
	\parbox{5in}{\footnotesize\smalllineskip Fig.~\thefigure. #1}
            \end{center}}
        \else
             {\begin{center}
	     {\footnotesize Fig.~\thefigure. #1}
              \end{center}}
        \fi}

\newcommand{\tcaption}[1]{
        \refstepcounter{table}
	\setbox\@tempboxa = \hbox{\footnotesize Table~\thetable. #1}
        \ifdim \wd\@tempboxa > 5in
           {\begin{center}
         \parbox{5in}{\footnotesize\smalllineskip Table~\thetable. #1}
            \end{center}}
        \else
             {\begin{center}
	     {\footnotesize Table~\thetable. #1}
              \end{center}}
        \fi}

\def\@citex[#1]#2{\if@filesw\immediate\write\@auxout
	{\string\citation{#2}}\fi
\def\@citea{}\@cite{\@for\@citeb:=#2\do
	{\@citea\def\@citea{,}\@ifundefined
	{b@\@citeb}{{\bf ?}\@warning
	{Citation `\@citeb' on page \thepage \space undefined}}
	{\csname b@\@citeb\endcsname}}}{#1}}

\newif\if@cghi
\def\cite{\@cghitrue\@ifnextchar [{\@tempswatrue
	\@citex}{\@tempswafalse\@citex[]}}
\def\citelow{\@cghifalse\@ifnextchar [{\@tempswatrue
	\@citex}{\@tempswafalse\@citex[]}}
\def\@cite#1#2{{$\null^{#1}$\if@tempswa\typeout
	{IJCGA warning: optional citation argument 
	ignored: `#2'} \fi}}

\def\pmb#1{\setbox0=\hbox{#1}
        \kern-.025em\copy0\kern-\wd0
        \kern.05em\copy0\kern-\wd0
        \kern-.025em\raise.0433em\box0}

\def\fnt#1#2{\footnotetext{\kern-.3em
        {$^{\mbox{\scriptsize #1}}$}{#2}}}

\def\fpage#1{\begingroup
\voffset=.3in
\thispagestyle{empty}\begin{table}[b]\centerline{\footnotesize #1}
        \end{table}\endgroup}

\def\runninghead#1#2{\pagestyle{myheadings}
\markboth{{\protect\footnotesize\it{\quad #1}}\hfill}
{\hfill{\protect\footnotesize\it{#2\quad}}}}
\font\tenbf=cmbx10
\font\tenit=cmti10 
\font\tenit=cmti10
\font\bfit=cmbxti10 at 10pt

\font\ninerm=cmr9

\font\eightrm=cmr8

\newlength{\mywidth}\mywidth=6cm 
\def\fref#1{fig.\ref{#1}}

\renewenvironment{figure}{%
    \refstepcounter{figure}
    }
{}
\def\fignum{{\bf Fig.\arabic{figure}.\quad}}

\def\lsym{\raise-3pt\hbox{\vbox{\tabskip0pt\offinterlineskip
	\halign{\tabskip0pt plus 1em
	##\tabskip0pt\cr
	$\,\,<\,\,$\cr
	$\,\,\sim\,\,$\cr}}}}
\def\rsym{\raise-3pt\hbox{\vbox{\tabskip0pt\offinterlineskip
     \halign{\tabskip0pt plus 1em
      ##\tabskip0pt\cr
      $\,\,>\,\,$\cr
      $\,\,\sim\,\,$\cr}}}}
\def\qed{\hbox{${\vcenter{\vbox{			
	\hrule height 0.4pt\hbox{\vrule width 0.4pt height 6pt
	\kern5pt\vrule width 0.4pt}\hrule height 0.4pt}}}$}}
\def\theequation{\thesection.\arabic{equation}}		

\begin{document}

\def\nn{\nonumber}
\def\df{\partial}
\def\half{{\textstyle{{1}\over{2}}}}

\runninghead{I.N.Nikitin and J.De Luca}
{Wheeler-Feynman electrodynamics}

\normalsize\textlineskip
\thispagestyle{empty}
\setcounter{page}{1}

~

\vspace{-3cm}

\vspace*{0.88truein}

\fpage{1}
\centerline{\bf NUMERICAL METHODS FOR THE 3-DIMENSIONAL 2-BODY PROBLEM} 
\vspace*{0.035truein}
\centerline{\bf IN THE ACTION-AT-A-DISTANCE ELECTRODYNAMICS} 
\vspace*{0.37truein}
\centerline{\footnotesize I.N.NIKITIN* and J. DE LUCA**} 
\vspace*{0.15truein}
\centerline{\footnotesize\it * German National Research Center 
for Information Technology,}
\vspace*{0.015truein}
\centerline{\footnotesize\it 53754 Sankt Augustin, Germany}
\vspace*{0.015truein}
\centerline{\footnotesize\it ** Departamento de F\'isica, Universidade Federal de S\~ao Carlos, Via Washington Luis, km 235,}
\vspace*{0.015truein}
\centerline{\footnotesize\it 13565-905 S\~ao Carlos, SP, Brazil}
\vspace*{0.15truein}
\centerline{\footnotesize E-mail: 
Igor.Nikitin@gmd.de $|$ deluca@df.ufscar.br  }

\vspace*{0.225truein}
\pub{20-03-2001}

\vspace*{0.21truein}
\abstracts{We develop two numerical methods
to solve the differential equations with deviating arguments
for the motion of two charges
in the action-at-a-distance electrodynamics. Our first method uses St\"urmer's
extrapolation formula and assumes that a step of integration can be taken as a 
step of light ladder, which limits its use to shallow energies. The second 
method is an improvement of pre-existing iterative schemes, designed for 
stronger convergence and can be used at high-energies. 
}{}{}

\vspace*{10pt}


\vspace*{1pt}\textlineskip	

\section{Action-at-a-distance electrodynamics}\label{sec1}
\vspace*{-0.5pt}
\noindent
In this paper we continue an investigation initiated in \cite{fw_jmpc0} and here extended to the numerical analysis of the 3-dimensional two-body problem in the
 relativistic action-at-a-distance electrodynamics of Wheeler and Feynman\cite{Fey-Whe} , henceforth called 3D WF. The dynamics follows from the Schwarzschild-Tetrode-Fokker\cite{Schw-Tetr-Fokk} direct-interaction functional. Equations of motion are derived from Hamilton's principle for the action integral 
\[
S=-%
\mathop{\textstyle\sum}%
_{i}%
\textstyle\int%
m_{i}cds_{i}-%
\mathop{\textstyle\sum}%
\limits_{i>j}(e_{i}e_{j}/c)%
\textstyle\int%
\textstyle\int%
\delta \left( \left\| x_{i}-x_{j}\right\| ^{2}\right) \dot x_{i}\cdot
\dot x_{j}ds_{i}ds_{j}, 
\]
where the four-vector $x_{i}(s_{i})$ represents the four-position of
particle $i$ parametrized by arc-length $s_{i}$, double bars indicate
four-vector modulus $\left\| x_{i}-x_{j}\right\| ^{2}\equiv
(x_{i}-x_{j})\cdot (x_{i}-x_{j})$ and the dot indicates Minkowski
relativistic scalar product of four-vectors with metric tensor
$g_{\mu\nu}=diag(1,-1,-1,-1)$. Integration is to be carried
over the whole particle trajectories, at least formally \cite{Starusk}, 
unlike other usual additive cases of classical mechanics. 
The above action
integral describes an interaction at the advanced and retarded light-cones
with an electromagnetic potential given by half the sum of the advanced and
retarded Li\`{e}nard-Wierchert potentials \cite{Anderson}. Wheeler and
Feynman showed that electromagnetic phenomena can be described by this
direct action-at-a-distance theory in complete agreement with Maxwell's
theory as far as the classical experimental consequences\cite{Fey-Whe,Leiter}%
. This direct-interaction formulation of electrodynamics was developed to
avoid the complications of divergent self-interaction, as there is no
self-interaction in this theory, and also to eliminate the infinite number
of field degrees of freedom of Maxwell's theory\cite{Plass}. It was a great
inspiration of Wheeler and Feynman in 1945, that followed a lead of 
Tetrode\cite{Schw-Tetr-Fokk} and showed that with the extra hypothesis 
that the electron
interacts with a completely absorbing universe, the advanced response of
this universe to the electron's retarded field arrives {\it at the present
time of the electron} and is equivalent to the local instantaneous
self-interaction of the Lorentz-Dirac theory\cite{Dirac} plus the retarded 
interaction among charges\cite{Narlikar}. The
action-at-a-distance theory is also symmetric under time reversal, as the
Fokker action includes both advanced and retarded interactions. Dissipation
in this time-reversible theory is due to interaction with the other charges
of the universe and becomes a matter of statistical mechanics of
absorption\cite{Einstein}. The area of Wheeler-Feynman electrodynamics has
been progressing slowly but steadily since 1945: quantization was achieved
by use of the Feynman path integral technique and the effect of spontaneous
emission was successfully described in terms of interaction with the future
absorber, in agreement with quantum electrodynamics. It was
also shown that it is possible to avoid the usual divergencies associated
with quantum electrodynamics by use of proper cosmological boundary
conditions\cite{Narlikar}. As far as understanding of the dynamics governed
by the equations of motion, the state of the art is as follows: the exact
circular orbit solution to the attractive two-body problem was proposed in
1946 by Schonberg\cite{Schonberg} and rediscovered in 1962 by 
Schild\cite{Schild}. The
1-dimensional symmetric two-electron scattering is a special case where the
equations of motion simplify a lot and it has been studied by many authors,
both analytically and numerically \cite{VonBaeyer,Driver1,Igor}. In this
very special case the initial value functional problem surprisingly requires
much less than an arbitrary initial function to determine a solution
manifold with the extra condition of bounded manifold for all times. It was
shown that the solution is uniquely determined by the inter-electronic
distance at the turning point if this distance is large enough\cite{Driver1}.
As a result of this theorem, there is a single continuous parameter 
(the positive energy) describing the unique non-runaway P-symmetric solution
at that given positive energy. Numerical analysis\cite{fw_jmpc0} shows that
at some critical energy this unique solution splits into 
three solutions, from which two are not P-symmetric themselves, 
but are P-conjugated to each other.

For the 3D WF the only known result is the linear stability analysis of the
Schonberg-Schild circular orbits\cite{Hans}, exhibiting an infinite number
of unstable solutions for the characteristic equation.
This paradoxical linear instability of circular orbits
is a warning that the first relativistic correction
can not be the generic unfolding of this complex dynamics
considered as a perturbation of the integrable Coulombian system. 
The question whether these unstable solutions are present in
the exact theory can be investigated numerically, which is a use for 
our integrators.

In the following we discuss three numerical methods for the
solution of 3D WF. The first one, originally 
proposed in \cite{Harvard}, involves solving the equations 
with respect to the most advanced velocities and accelerations, 
converting the equation of motion to a double-delay form. 
As a result, the evolution becomes completely 
and quite naturally determined by the past trajectory, 
allowing an easy numerical implementation. 
However, in section 2.1 
we show that this numerical scheme is largely unstable 
and can not be used in practice. We discuss the reasons
of this instability, which are closely related to the existence of
unstable solutions of the characteristic equation\cite{Hans}.
In section 2.2 
we describe a method, based on usage of special parametrization of world lines
(light-ladder parametrization \cite{fw_jmpc0}), which implements the 
direct integration of the equations when the integration step is equal to one 
step of light ladder. The method is applicable at low velocities
$v/c\sim10^{-2}$ and is stable at least for integration times up to $N\sim10^{6}$ revolutions of the particles. 
In section 2.3 
we consider an iterative method, originally proposed in \cite{VonBaeyer}
to solve the 1-dimensional 2-body WF problem, which after
certain modifications can be applied to the 3-dimensional case.
This method works with shorter intervals of integration than the light-ladder 
and it is able to resolve the structure of solutions for intervals smaller 
than one light ladder step and also converges at high energies. The results 
obtained using the methods are presented in section 3. 

The mathematical structure of the equations of motion of 3D WF is described
in Appendix~A. Poincare's invariance of the Fokker Lagrangian 
implies the associated non-local Noether's constants of motion as integrals 
over a segment of the past history\cite{Anderson,Fey-Whe,Schild}. 
These conserved integrals provide 
a useful check for the numerical work and we discuss them as well 
in Appendix~A. Further appendices provide necessary details
about numerical schemes.  

\section{Computational methods}

\vspace{-5mm}
\subsection{Solving the equations with respect 
to the advanced variables}\label{sec2a}

The most intuitively attractive method, which is based on solving 
the equations of motion for the most advanced velocities and accelerations,
turns out to be numerically unstable. The reason
for this instability can be easily understood considering the simpler
1-dimensional version of the problem with repulsive potential. We henceforth use the letter $x$ to denote the cartesian vector of position for particle 1 and the letter $y$ to denote the position of particle 2. The velocity and acceleration  evaluated at the present time of particle 1 are written as $v_x$ and $a_x$ respectively. An upper plus (or minus) above quantities signifies that the quantity is evaluated at the future (or past) point of the light cone starting from the present position of other particle. The equations 
of motion for 1-dimensional repulsive motion have the simple form
$$a_{x}=\half(1-v_{x}^{2})^{3/2} 
\left({1+v_{y}^{-}\over1-v_{y}^{-}}\cdot(y^{-}-x)^{-2} 
+ {1-v_{y}^{+}\over1+v_{y}^{+}}\cdot(y^{+}-x)^{-2}\right),$$
in units $e=m=c=1$. The equations for the other particle are obtained by exchanging $x\leftrightarrow y$ in the above formula.
The idea is to solve the above equation of motion of particle $x$ for the most 
advanced velocity $v_{y}^{+}$ of particle $y$ as
$$
v_{y}^{+}={{1-G}\over{1+G}},\quad
G\equiv (y^{+}-x)^{2}\cdot\left[{2a_{x}\over(1-v_{x}^{2})^{3/2}}
-{1+v_{y}^{-}\over1-v_{y}^{-}}\cdot(y^{-}-x)^{-2}\right].
$$ 
The equation for $v_{x}^{+}$ is again obtained by interchanging 
$x\leftrightarrow y$. It is seen that we have a functional differential 
equation with delayed argument only, as first pointed out in\cite{Harvard}.
As far as initial conditions go, the general theory on delay equations \cite
{Elsgolts} tells us that we need to provide an initial $C^{2}$ function
describing the position of particle $x$ in the past, as well as the 
information on particle $y$ needed is also to be provided 
over twice the retardation lag seen by particle $y$.

The first difficulty with such setting of the problem
is that the expression for $v_{y}^{+}$ includes the {\it derivative}
$a_{x}=dv_{x}/dt_{x}$ evaluated in the retarded time.
In the numerical solution this derivative should be estimated
from known values $v_{x}(t_{x}^{k})$ on the integration grid
using an appropriate finite difference scheme.
The common feature of these schemes is that they contain
a factor $h^{-1}$, where $h$ is the integration step.
Supposing that points $v_{x}(t_{x}^{k})$ on the grid
have a computational error of order $\epsilon$, 
we will have an error in the derivative
$dv_{x}/dt_{x}$ of order $\epsilon h^{-1}$. Due to the functional
relationship $v_{y}^{+}=f(dv_{x}/dt_{x})$, 
$v_{y}^{+}$ will have the error $\epsilon h^{-1}$. We see that the 
numerical scheme amplifies the computational error by a factor of $h^{-1}$ 
after each step of light ladder. The integration step for the given equation
should be small, typically $h<10^{-2}$, and as a result,
the scheme works only for 2-3 light ladder steps and then explodes.

The second difficulty concerns the possibility of continuation 
of the trajectory.
It turns out that the advanced velocity given by the above expression
is not guaranteed to be bounded by the velocity of light.
Even in the case where numerical integration would be possible,
for some initial data, at a certain point $v_{y}^{+}$ can exceed 
the velocity of light and the solution cannot be continued to the next 
light ladder step (where this $v_{y}>1$ should enter under the square root
$(1-v_{y}^{2})^{1/2}$). At least for symmetric 1-dimensional orbits 
Driver's theorem\cite{Driver1} says that the set of infinitely continuable
solutions is finite-dimensional, implying that for most initial functions
the solution can not be continued for all times, 
and only a thin finite-dimensional subset of initial functions produces
the good solutions. Ideally, in the given approach 
one should find special rare initial data, for which the solution can be 
continued for a large time, and this is easily seen to be a formidable 
numerical task. 

These same difficulties plague the 2- and 3-dimensional cases, with
the extra complication that the equations of motion include the advanced
acceleration in a degenerate way. An algebraic constraint appears 
and we solve it in Appendix~B. As a result of this constraint, 
one component of the advanced velocity can be found from the previous acceleration 
(and other variables) by a functional relation of the form $v^{+}=f(a,...)$.
This property again makes the resulting numerical scheme unstable, 
like in the 1-dimensional case.

\subsection{Direct integration scheme (low energy)}\label{sec2b}

In the limit of small velocities a step of light ladder 
is small and can be taken as an integration step.
The following computational scheme can work in this limit:

\begin{figure}\label{scheme}
\begin{center}

\epsfysize=3cm\epsffile{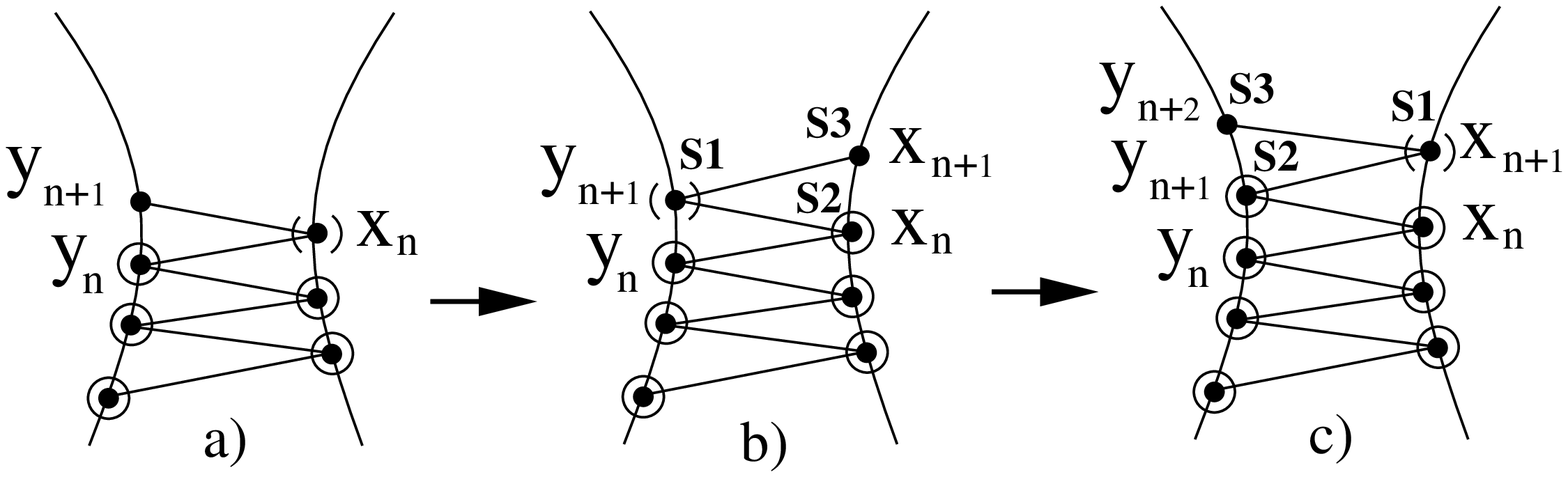}

\fignum Direct integration scheme.

\end{center}
\end{figure}

Suppose that coordinates and velocities are known in points
$x_{k},k\leq n;\ y_{k},k\leq n+1$ (shown by dots on \fref{scheme}a), 
and accelerations are known (stored from past integration)
in points $x_{k},k\leq n-1;\ y_{k},k\leq n$ (shown by circles). 
We then go through the following steps:

\noindent S1. Because the step of light ladder is small,
we are able to compute the acceleration in point $y_{n+1}$ 
as the left derivative, using a finite difference scheme, 
described further. This computed derivative is shown by brackets in
\fref{scheme}b.

\noindent S2. As a result, we are able to find 
the acceleration in point $x_{n}$, using the equations of motion.
This is shown in \fref{scheme}b by a circle around point $x_{n}$.

\noindent S3. We are then able to perform one integration step, applying 
St\"urmer's method (described further), and find coordinates
and velocities in point $x_{n+1}$.

Next we apply these three steps for $x\leftrightarrow y$ interchanged.
As a result, we have \fref{scheme}c with the same pattern 
of data distribution, as in \fref{scheme}a, only with shifted index $n\to n+1$.

\noindent{\it Note:} during this process we actually have 
two definitions of acceleration in each point: computed by current 
and past velocities as left derivative (shown by brackets),
and computed from the equations of motion (shown by circles). These two
accelerations should coincide, and their relative difference 
$|\Delta\vec a|/|\vec a|$ can be used to check the consistency of the method
(see Appendix~C).

\paragraph*{St\"urmer's integration method} for the equation $\dot x=f(t,x)$
is described in \cite{Elsgolts},p.20. This $n$th order formula 
can be obtained from the following Taylor expansions:
\begin{eqnarray}
&&q_{k}=hf(t_{k},x_{k}),\nn\\
&&q_{k-1}=q_{k}-\ddot x_{k}h^{2}
+{\textstyle{1\over2}}x_{k}^{(3)}h^{3}-...+o(h^{n}),\label{Stursys}\\
&&q_{k-2}=q_{k}-2\ddot x_{k}h^{2}
+2x_{k}^{(3)}h^{3}-...+o(h^{n}),\nn\\
&&q_{k-n+1}=q_{k}-(n-1)\ddot x_{k}h^{2}
+{\textstyle{(n-1)^{2}\over2}}x_{k}^{(3)}h^{3}-...+o(h^{n}),\nn
\end{eqnarray}
by solving these $(n-1)$ linear equations for the
$(n-1)$ unknowns $x_{k}^{(p)}h^{p}$ and substituting them into the Taylor 
series for $x_{k+1}$
\begin{eqnarray}
&&x_{k+1}=x_{k}+q_{k}+{\textstyle{1\over2}}\ddot x_{k}h^{2}+
{\textstyle{1\over6}}x_{k}^{(3)}h^{3}+...+o(h^{n}).\nn
\end{eqnarray}
Particularly, for $n=8$ we have 
\begin{eqnarray}
&&x_{k+1}=x_{k}+q_{k}+{\textstyle{1\over2}}\Delta q_{k-1}
+{\textstyle{5\over12}}\Delta^{2}q_{k-2}
+{\textstyle{3\over8}}\Delta^{3}q_{k-3}
+{\textstyle{251\over720}}\Delta^{4}q_{k-4}\nn\\
&&+{\textstyle{95\over288}}\Delta^{5}q_{k-5}
+{\textstyle{19087\over60480}}\Delta^{6}q_{k-6}
+{\textstyle{5257\over17280}}\Delta^{7}q_{k-7}+o(h^{8}),\label{Stur8}
\end{eqnarray}
where $\Delta^{p}q_{k}$ are defined recurrently as 
$$\Delta q_{k-1}=q_{k}-q_{k-1},\ 
\Delta^{p}q_{k-1}=\Delta^{p-1}q_{k}-\Delta^{p-1}q_{k-1}.$$

The second derivative $\ddot x_{k}$, found from (\ref{Stursys})
at $n=8$, is given by the following backward differentiation formula:
\begin{eqnarray}
&&\ddot x_{k}=h^{-2}\sum\limits_{p=1}^{7}
{\textstyle{1\over p}}\Delta^{p}q_{k-p}+o(h^{6})\label{BDF}
\end{eqnarray}

\noindent
{\it Notes:}

\noindent 1. The advantage of St\"urmer's method is that the right hand sides 
of the equations do not need to be estimated in the intermediate points 
of the grid $x_{k}$, as for example with the type of Runge-Kutta methods,
thus the necessity of interpolation of data between 
integration points is avoided.

\noindent 2. Usage of lower order scheme reduces the precision
of integration, but improves stability of the method. 
Particularly, the reduction of order $n=8\ \to\ n=4$, 
correspondingly reduces the precision of conservation
of Noether's integrals (see appendices), and increases the 
upper velocity, for which the method is convergent 
$v=0.006\ \to\ v=0.02$.

\subsection{Iterative scheme (high energy)}\label{sec2c}

The following scheme, originally applied in \cite{VonBaeyer} to solve the
1D WF with repulsive interaction, is here improved and implemented to integrate the 3D WF in the highly relativistic limit. The method works as follows:

\vspace{2mm}\noindent\parbox{8cm}{
An initial trajectory is guessed for the two particles, 
which is called iteration number zero. We then use the equations 
of motion to evaluate the present force caused by the past data points 
$x^{-},y^{-}$, located on current trajectory, and future data points 
$x^{+},y^{+}$, {\it taken on the trajectory of previous iteration.} 
This acceleration is used as a force field to integrate the trajectory, 
producing the next iterated trajectory.
Repeating such integration of trajectories for fixed data 
on the initial segment, we arrive (if the iterations converge) 
to a solution of the WF problem for the initial data. 
As mentioned in \cite{VonBaeyer}, this scheme is not guaranteed 
to be convergent, and we developed special methods of stabilization 
(see Appendix~C), 
that is able to keep convergence up to very high values of the velocity. 
}\quad\quad\parbox{3.5cm}{\begin{figure}\label{iter}
~\epsfxsize=3cm\epsffile{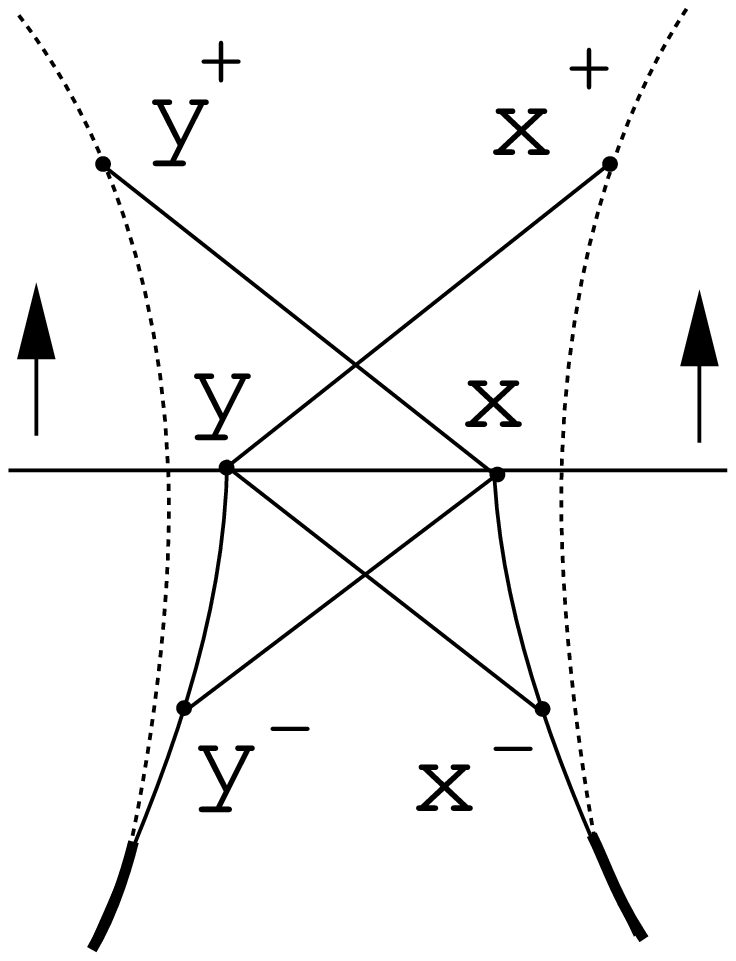}
\hbox{\fignum~Iterative~scheme.}
\end{figure}}

\noindent
We start our iterations from the analytically known 
circular solutions\cite{Schild}, fixing the initial segments 
of trajectories $\tau\in[0,0.5]$ to arcs of circular orbits 
through all iterations. Then, to obtain non-circular orbits, 
we apply short pulses of external force to both particles 
immediately after the segment of initial setting. 
The method allows to subdivide each light ladder step into a large number
of integration steps of sufficiently small size to have the correct integration
at high energies. St\"urmer's method, described in the previous section, 
can be used for integration. Special boundary conditions 
should be used at the end of integration interval, where no information 
is available about the advanced Lorentz force. These and further 
details about the method can be found in Appendix~C.

This iterative method requires storage of the complete integrated evolution 
from the current and previous iterations.  Due to memory restrictions, only
a short part of the trajectories (up to 10 revolutions) can be determined
by this method. Nevertheless, the method can provide useful information 
about structure of solutions at high energy when only short term
evolution is needed.

\section{Results}\label{sec3}

The computational methods, described in the previous sections, are applicable 
to arbitrary masses and charges. However, the case of equal masses
possesses additional symmetries, and here we present the results obtained 
for equal masses and opposite charges (positronium-like system).
Following Andersen and von Baeyer \cite{Hans}  
we fix the unit system to $c=1$, $e_{x}=-e_{y}=1$, $m_{x}=m_{y}=2$, 
so that reduced mass $m=(m_{x}^{-1}+m_{y}^{-1})^{-1}=1$. 
We present the evolution of the system in the center-of-mass frame (CMF),
using Schild's definition of CMF origin, 
given by Eq.(2.13) in\cite{Schild}.

\begin{figure}\label{f0}
\begin{center}

~a)\epsfysize=4.5cm\epsffile{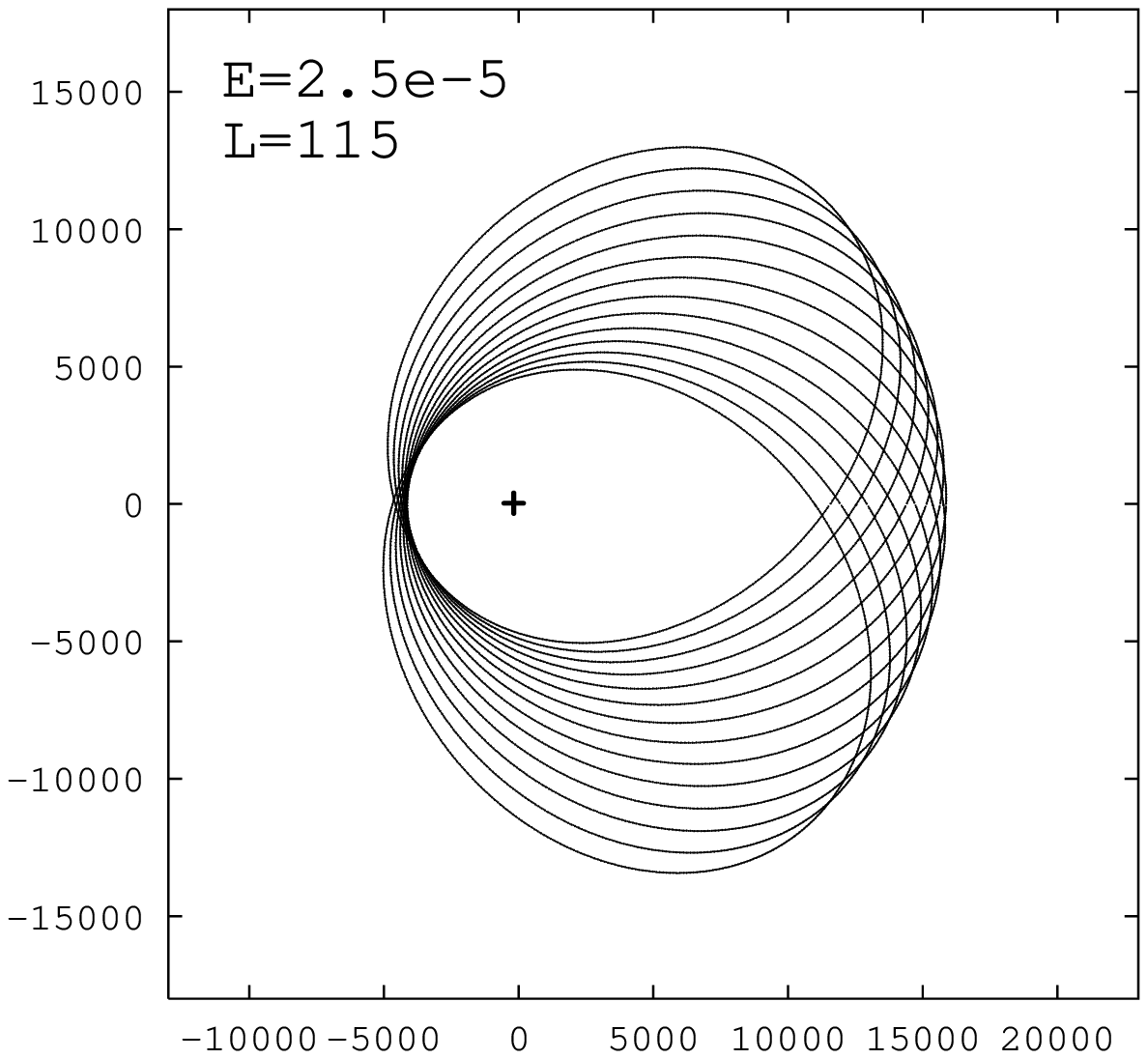}
~b)\epsfysize=4.5cm\epsffile{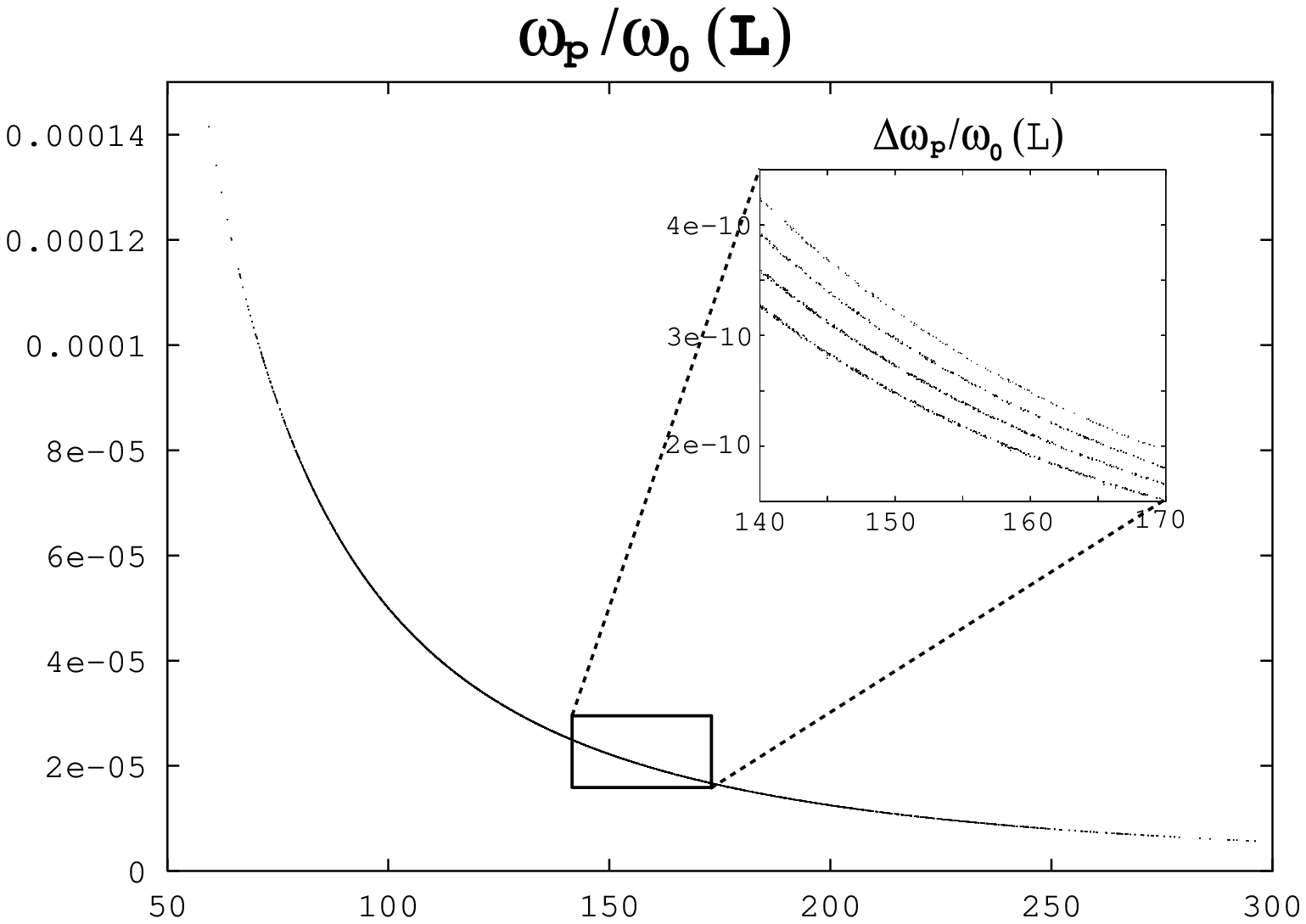}

~c)\epsfysize=4cm\epsffile{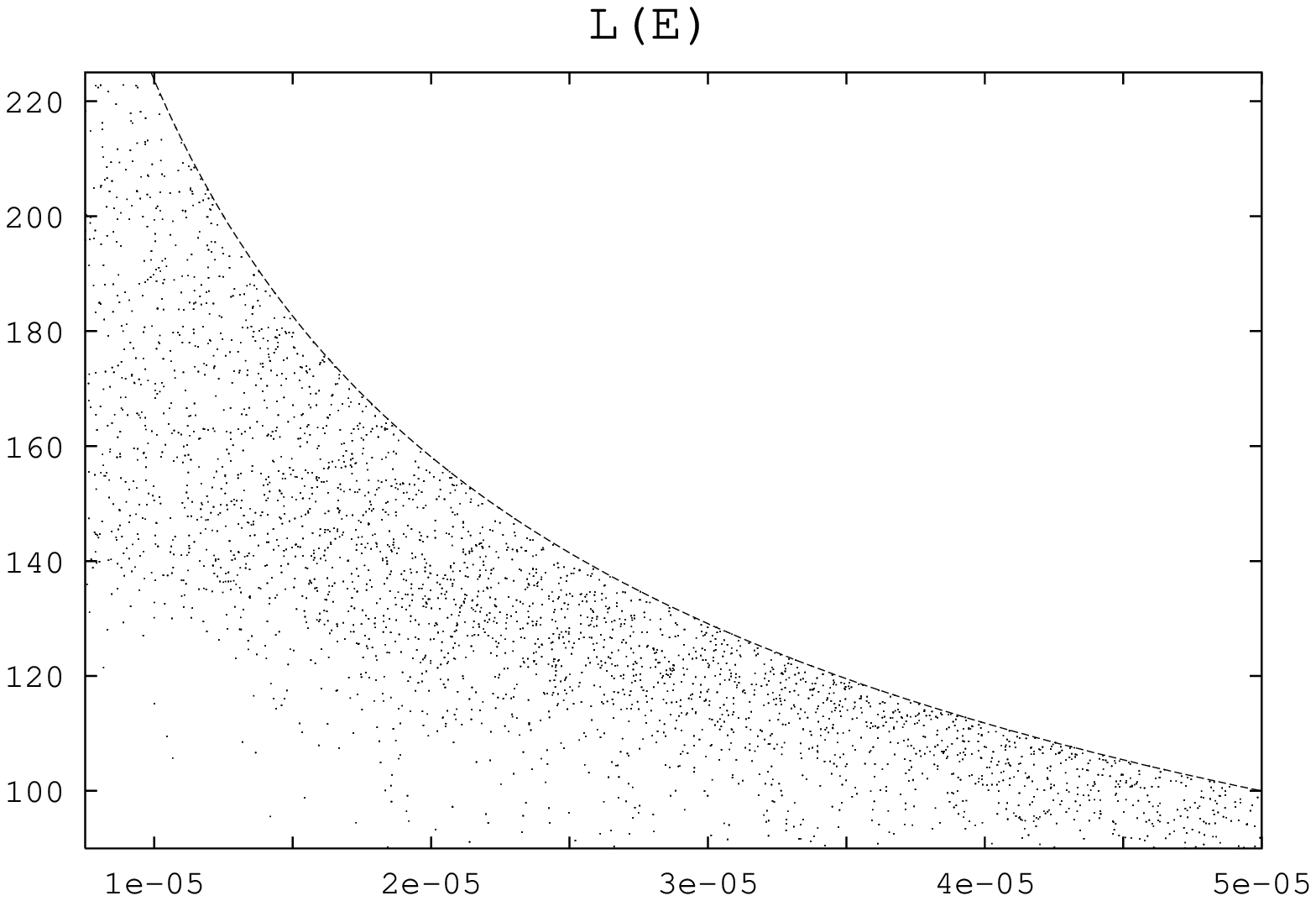}
~d)\epsfysize=4cm\epsffile{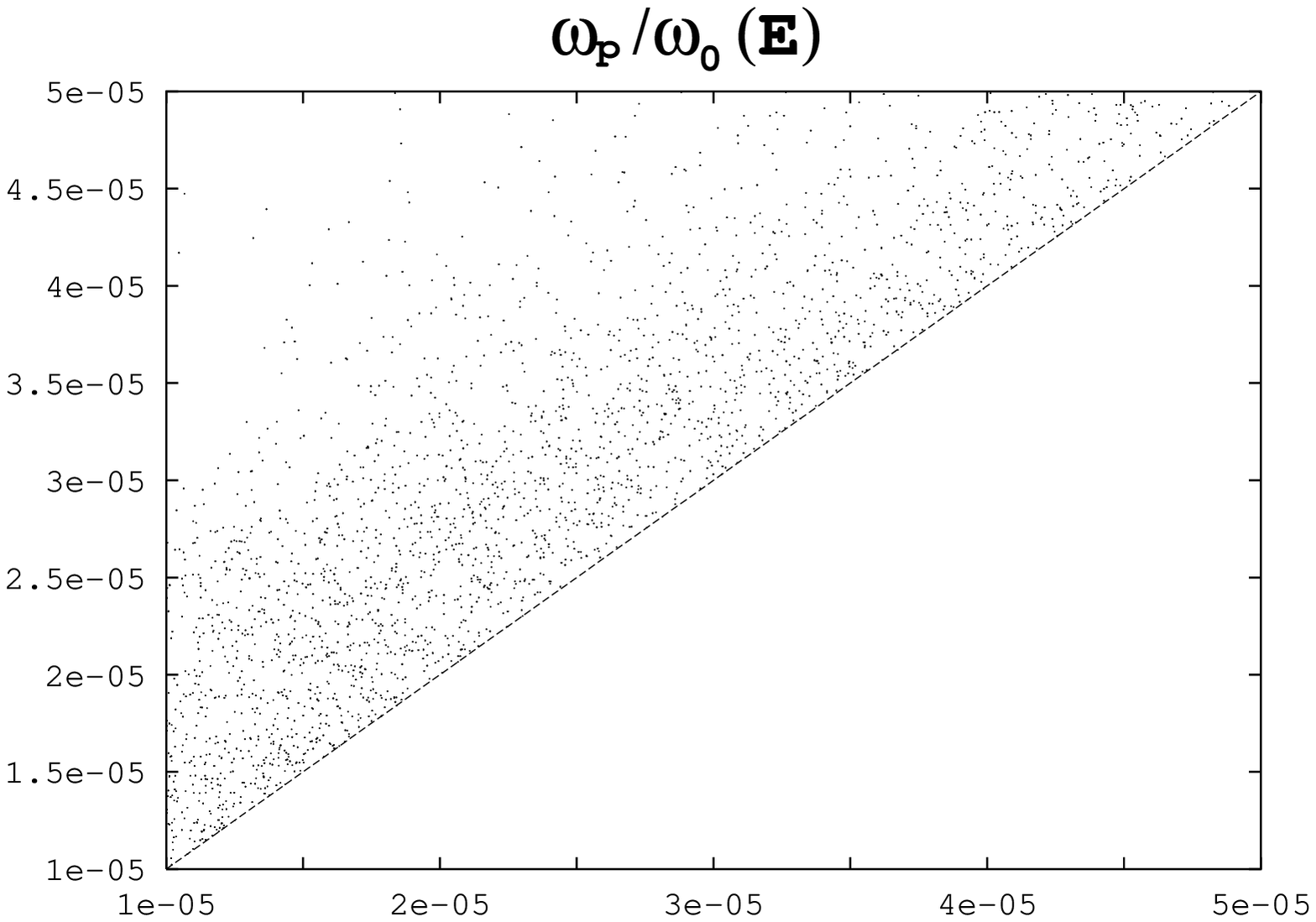}

\fignum Low-energy solutions: a) trajectories, 
b-d) related parameters.

\end{center}
\end{figure}

\paragraph*{Low energy: Darwin's precession.} 
At low energy the solution looks like very slowly precessing ellipse, 
shown on \fref{f0}a. Between each two consequential ellipses of this figure
500 intermediate ellipses are not shown. The figure presents 
two coincident curves: trajectory of particle $x$ and 
trajectory of particle $y$, reflected about the origin in CMF,
demonstrating the P-symmetry of the solution. 

\paragraph*{Related parameters} $(E,L,\omega_{0},\omega_{p})$:
binding energy $E=2m-\sqrt{P^{2}}$, angular momentum in the CMF, 
angular frequencies of rotation and precession. For 10,000 solutions 
generated by the methods described in Appendix~C, 
we measure these four parameters, and analyze the
4-dimensional scatter plot. Figures \ref{f0}c,d represent 
projections of the scatter plot onto planes $(E,L)$ 
and $(E,\omega_{p}/\omega_{0})$. The boundary curves on these figures
correspond to the circular orbit limit, 
and are given by formulae Eqs.(3.4),(3.5) in \cite{Schild}
and Fig.4, Eq.(5.1) in \cite{Hans}, which for the case of equal masses 
and low velocities reduces to 
$E=1/(2L^{2})=\omega_{p}/\omega_{0}$.

Figure \ref{f0}b represents a diagram $(L,\omega_{p}/\omega_{0})$,
where the scatter plot is projected almost to a curve.
It shows that the ratio $\omega_{p}/\omega_{0}$ depends mainly on $L$
as $\omega_{p}/\omega_{0}=1/(2L^{2})$. This dependence was
predicted by Darwin \cite{Darwin}, who first studied the lowest order
relativistic corrections to the Coulomb problem and found the
described effect of precession. Using non-perturbative methods,
we detect deviations from Darwin's law by an order $10^{-10}$, 
still resolvable by our measurement (which gives absolute precision of
$\omega_{p}/\omega_{0}$ about $10^{-14}$). Inner image shows
the difference $\Delta\omega_{p}/\omega_{0}=
1/(2L^{2})-\omega_{p}/\omega_{0}$ versus $L$, 
for 4 narrow bands $2EL^{2}=~[0.98,1],~[0.78,0.8],~[0.58,0.6],~[0.38,0.4]$,
following on the image in the order down-up. Thus, $\omega_{p}/\omega_{0}$ 
depends mainly on $L$ and very weakly on $E$.

The structure of the graphs shows that the scatter plot
in 4-dimensional space $(E,L,\omega_{0},\omega_{p})$ is located on
a 2-dimensional surface, like in the non-relativistic Coulombian problem,
where $(E,L)$ are the only parameters, defining the shape of solutions 
in the CMF. Another similarity with Coulombian problem is that
the trajectories in CMF with 
high precision\footnote{Motion along $\vec L$-axis 
consists of small oscillations with amplitude $\Delta z=10^{-8}$ 
of classical radius or relative value 
$\Delta z/(\mbox{XY-size of orbit})=10^{-12};$ 
and slow Brownian motion with the velocity about $v=10^{-17}$; 
both motions have the same order as non-conservation of 
Noether's integrals, appearing due to computational errors.} 
~belong to a plane, perpendicular to the vector of angular momentum $\vec L$,
even if initial data in laboratory frame are non-planar.
In relativistic 3D WF problem we cannot prove planarity of orbits
so easily as in Coulombian problem, and currently we have it only 
as a result of numerical experiment. 

\begin{figure}\label{xy}
\begin{center}
~\epsfysize=5cm\epsffile{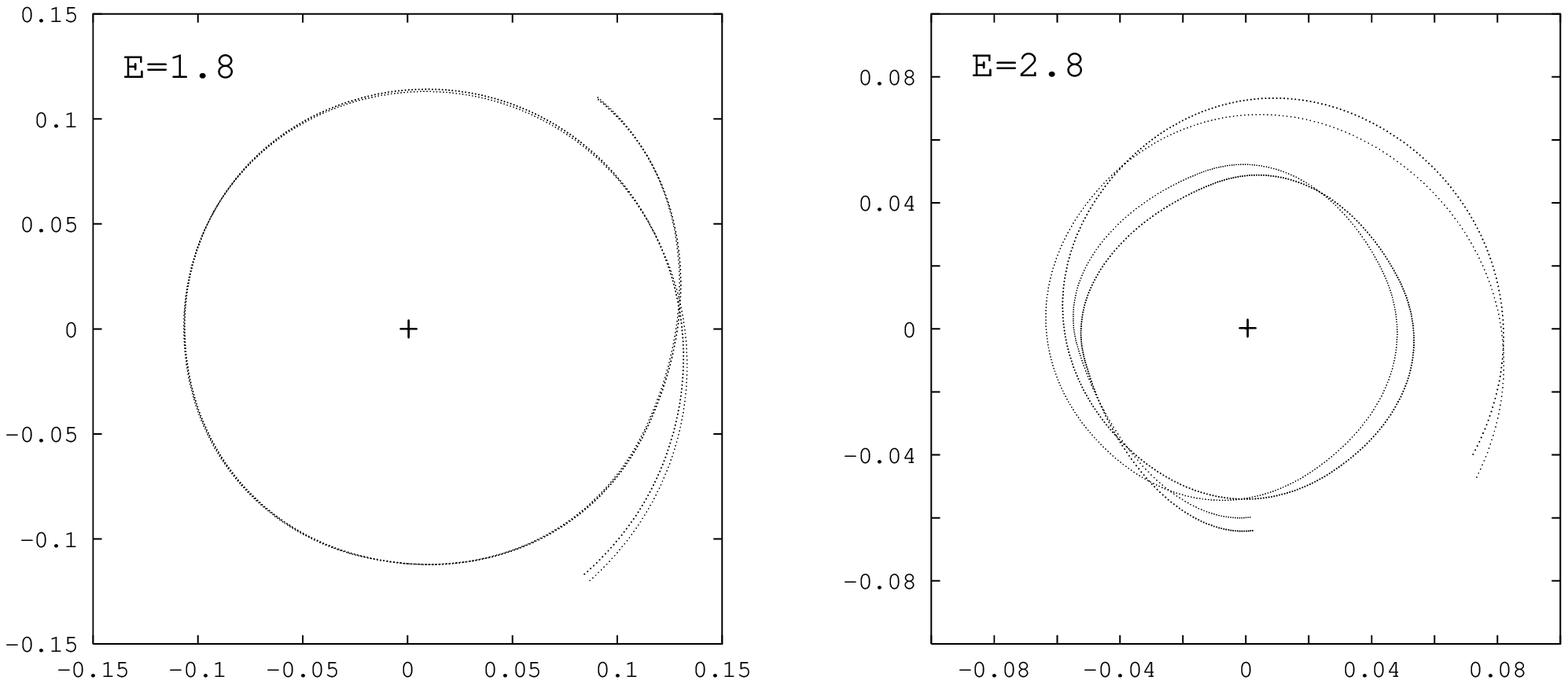}

\fignum High-energy solutions: trajectories.
\end{center}
\end{figure}

\paragraph*{High energy: Bifurcations.} Figure \ref{xy}
shows pieces of the high-energetic trajectories, found by method 
sec.2.3. Here we again superpose trajectories $x$ and $(-y)$ to test the 
P-symmetry of the solution. On the left image a small asymmetry is visible at the beginning 
of the trajectories, caused by the asymmetrical statement of the problem 
at the beginning (see Appendix~C), which exponentially disappears 
in the inner regions of the trajectories. When the energy increases, 
the asymmetry penetrates deeper to inner parts of the trajectory, 
and at high energy the solution completely looses the symmetry.

\begin{figure}\label{xy1}
\begin{center}
~\epsfysize=5cm\epsffile{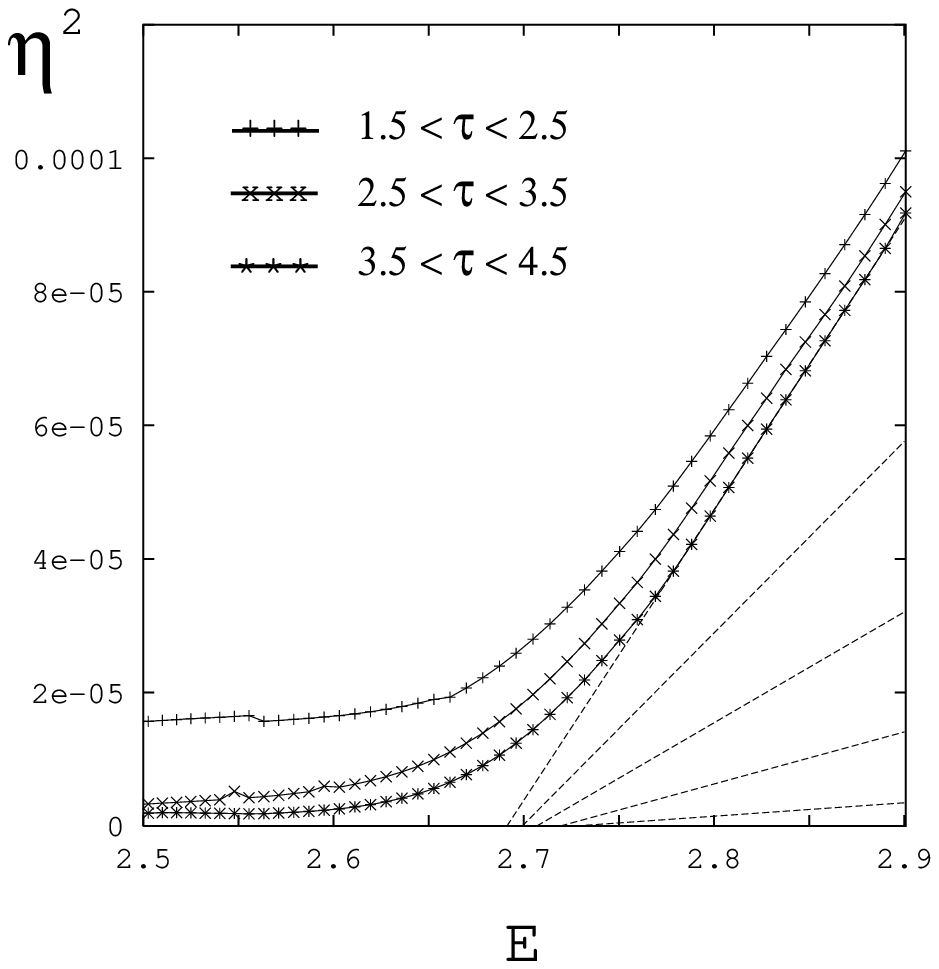}
~\epsfysize=5.6cm\epsffile{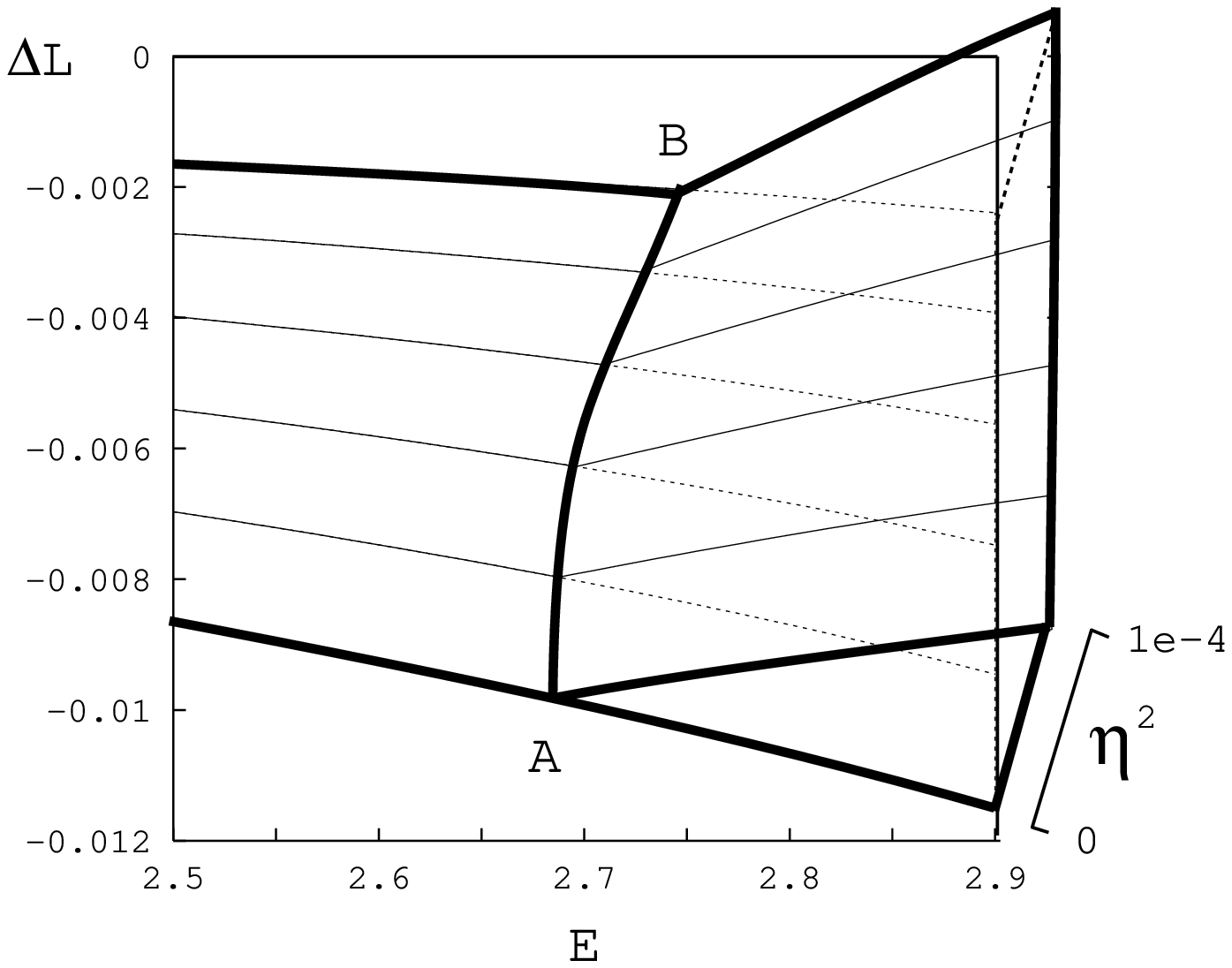}

\fignum High-energy solutions: related parameters.
\end{center}
\end{figure}

We define the measure of asymmetry as $\eta=\max|\vec x(t)+\vec y(t)|$
in CMF, where the maximum is taken over a step of light ladder. 
Then we draw graphs $\eta^{2}(E)$ 
for consequential light ladder steps -- they are given by
curves with points on the left graph of \fref{xy1}. 
We see that at low energies 
the asymmetry tends to zero with increasing number of light ladder steps, 
while at high energies the asymmetry tends to a non-zero value.
The limiting $\eta^{2}$ is well described by a linear function of E
(corresponding to a supercritical pitchfork-type bifurcation 
$\eta\sim\sqrt{E-E_{0}}$),
it is shown by upper straight line on the graph. Lower straight lines
correspond to analogously constructed dependencies for smaller
asymmetry of initial conditions. Right image presents the result of assembling 
all pictures to 3D graph. On this graph $\Delta L(E)=L(E)-L_{0}(E)$,
$L_{0}(E)$ is angular momentum for circular orbits.
Line AB subdivides the diagram onto two parts:
{\it symmetric phase} -- on the left from AB the trajectories
possess P-symmetry and the shape of solution is uniquely 
defined by two parameters $(E,L)$; {\it asymmetric phase} -- 
on the right from AB the symmetry 
is violated and given $(E,L)$ correspond to a continuous set 
of solutions, differing by measure of asymmetry.

Such a structure of the phase space was actually predicted
by Andersen and von Baeyer\cite{Hans}, 
from the linear stability analysis of the circular orbits in 2D WF. 
This work shows that 
at velocities of about $v=0.95$, new real eigenvalues of
the characteristic equation appear, representing new degrees of
freedom in the system. Corresponding linear mode,
denoted by Andersen and von Baeyer as $D^{(-)}$, violates the
P-symmetry. In our non-perturbative approach we detect this effect
in the region of energy $E\in[2.6,2.8]$, which for circular orbits
corresponds to the region of velocities $v\in[0.937,0.954]$, 
in good agreement with the predictions \cite{Hans}. 
Actually, in point B, where the orbits are close to circular, 
we have $E=2.75,\ v=0.950$, i.e. exact coincidence.

\paragraph*{Conclusion.} Two numerical methods are developed
to solve 3D WF problem in the shallow and high energy ranges.
The methods go further than pre-existing ones\cite{Hans,Darwin},
based on first order perturbation theory. Our
non-perturbative analysis reproduces part of the
pre-existing results with the exception of the divergent
modes predicted in \cite{Hans}, whose exponential increase violates the
limitations of the linear perturbation theory\cite{Hans}. The question
about their existence in the complete theory is open. Our numerical
methods do not reproduce such solutions as far as the unforced equations
of motion are considered. On the other hand, some traces of divergent
modes are present in the solution of forced equations, describing the
reaction to an external perturbation. The results of this analysis
can be found in \cite{www} and will be published elsewhere.

\paragraph*{Acknowledgments.}
We would like to thank Dr. Martin G\"obel for the warm hospitality 
in GMD.IMK.VE, where this article was completed, and 
a grant from FAPESP that supported one of us in Germany (J. De Luca).

\baselineskip=0.4\normalbaselineskip\footnotesize

\paragraph*{Appendix A: equations of motion and conserved quantities} ~~~


\vspace{2mm}\noindent 1. 
The equations of motion for the 3D WF
have the following form:

\begin{eqnarray}
&&\vec a_{x}^{(\pm)}={{e_{x}}\over{m_{x}}}\sqrt{1-v_{x}^{2}}
\left(\vec E_{x}^{\pm}\left(1-{{(\vec r_{x}^{\pm}\vec v_{x})}
\over{r_{x0}^{\pm}}}\right)
+(\vec E_{x}^{\pm}\vec v_{x})\left({{\vec r_{x}^{\pm}}
\over{r_{x0}^{\pm}}}-\vec v_{x}\right)\right),\label{A1}\\
&&\vec E_{x}^{\pm}={(\mp)e_{y}\over(r_{x0}^{\pm}-
(\vec r_{x}^{\pm}\vec v_{y}^{\pm}))^{3}}(
(1-(\vec v_{y}^{\pm})^{2}-(\vec r_{x}^{\pm}\vec a_{y}^{\pm}))
(\vec r_{x}^{\pm}-r_{x0}^{\pm}\vec v_{y}^{\pm})
+(r_{x0}^{\pm}-(\vec r_{x}^{\pm}\vec v_{y}^{\pm}))
r_{x0}^{\pm}\vec a_{y}^{\pm}),\nn\\
&&\vec r_{x}^{\pm}=\vec y^{\pm}-\vec x,\quad 
r_{x0}^{\pm}=\pm|\vec r_{x}^{\pm}|,\quad
\vec a_{x}=\half(\vec a_{x}^{(+)}+\vec a_{x}^{(-)})+
{{1}\over{m_{x}}}\sqrt{1-v_{x}^{2}}
(\vec F_{ext}-(\vec F_{ext}\vec v_{x})\vec v_{x}).\nn
\end{eqnarray}
Here $\vec a_{x}$ denotes the acceleration of particle $x$;
variables $\vec E_{x}^{\pm}$ are advanced and retarded components
of electric field, created by particle $y$ at the location of particle $x$;
variables $\vec y^{\pm},\vec v_{y}^{\pm},\vec a_{y}^{\pm}$ 
stand correspondingly
for position, velocity and acceleration of particle $y$
at time moments, where the world line of particle $y$ is
intersected by the light cone with origin in $x$. Equations of motion  
for particle $y$ are given by replacement $(x\leftrightarrow y)$
in these expressions. The term $\vec F_{ext}$ 
corresponds to the force of external perturbation, used to
produce perturbed initial conditions starting from the exact 
circular solution. For the external force
we choose a cap-like form $\vec F_{ext}(t)=\vec F_{a} 
\exp[(1-(\Delta t/(t-t_{0}))^{2})^{-1}], 
\ \mbox{if}\ |t-t_{0}|<\Delta t;\ \mbox{otherwise}~0.$
This function is infinitely differentiable and has a finite support. 

\vspace{2mm}\noindent 2. 
In our approach the equations of motion
are integrated in the light ladder parametrization, which was 
introduced in \cite{fw_jmpc0} and modified by us in the following way:

\noindent\parbox{9cm}{
On the initial segments of the world lines, shown in bold on \fref{ladder},
we select arbitrary monotonous parametrization $\tau\in[0,0.5]$,
e.g. taking $\tau$ as a linear function of the time components $x_{0},y_{0}$.
We then consider light rays, which consequentially reflect 
between the world lines, and distribute the parametrization along these rays, 
adding $\tau\to\tau+0.5$ at the points of 
reflection\footnotemark.
In this parametrization $x(\tau)$ and $y(\tau\pm0.5)$ are related 
by light rays, so that deviation of the argument in the equations 
of motion becomes constant. As a result, while determining the 
intersection of the light cone with a world line, we are taking 
the data directly from one of the past integration points,
not solving the equation of intersection and not performing
high order interpolations of data between integration points.
}\quad\quad\parbox{3cm}{\begin{figure}\label{ladder}
~\epsfysize=3cm\epsffile{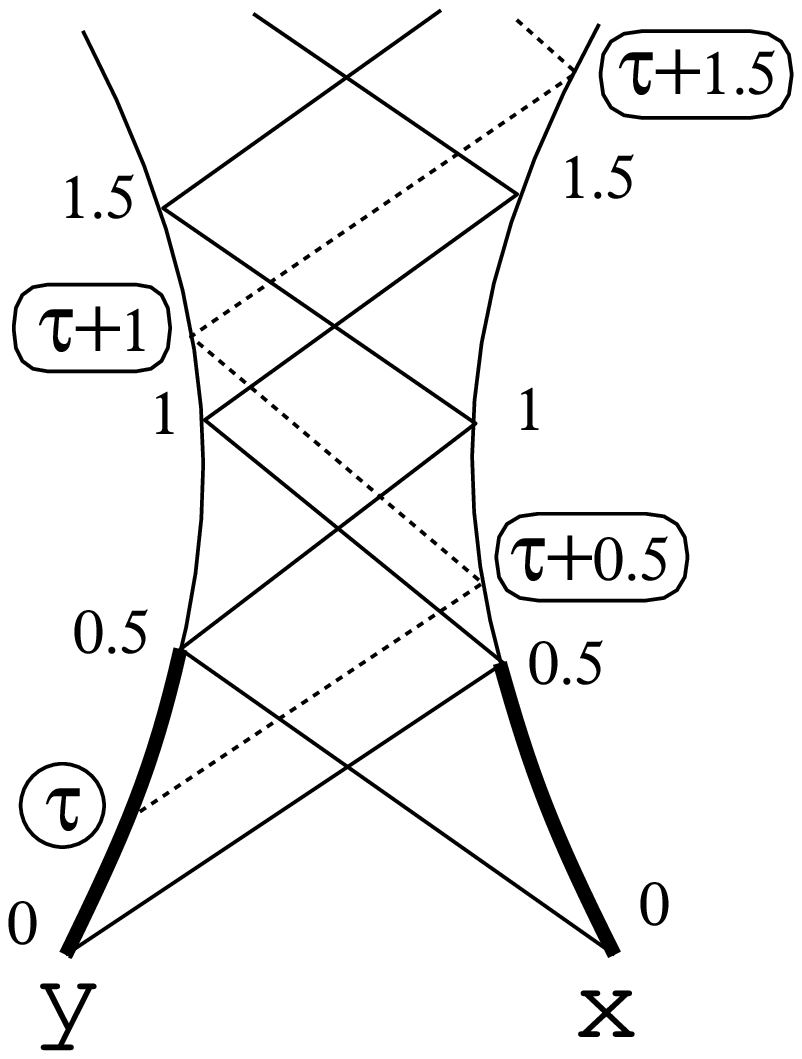}
\fignum Light~ladder parametrization.
\end{figure}}

\footnotetext{This procedure gives a $C^{\infty}$-parametrization 
under special conditions in the vicinity of the point $\tau=0.5$, 
which are satisfied particularly when the initial segments 
of the trajectories are set to circular orbits.}

Using the light ladder parametrization, we consider the evolution
of the variables $\vec x(\tau),\vec y(\tau),$ $x_{0}(\tau),y_{0}(\tau)$
reformulating the equations of motion in terms of
$\dot{\vec x}=\dot x_{0}\vec v_{x},\ \dot{\vec v_{x}}=\dot x_{0}\vec a_{x},$
and the analogous pair for the other particle $y$. 
We substitute for $\dot x_{0}$ a formula
$\dot x_{0}=\dot y_{0}^{-}(|\vec r_{x}^{-}|+(\vec r_{x}^{-}\vec v_{y}^{-}))/
(|\vec r_{x}^{-}|+(\vec r_{x}^{-}\vec v_{x}))$, which follows from
differentiation of the light cone condition $r_{x\mu}^{-}dr_{x\mu}^{-}=0$
and expresses $\dot x_{0}$ in terms of the past variables 
$(\vec y^{-},\vec v_{y}^{-},\dot y_{0}^{-})$ and current variables 
$(\vec x,\vec v_{x})$. For $\vec a_{x}$ we substitute the equations of
motion given above (variable $\vec a_{y}^{-}$, participating in
r.h.s. of the equations, is taken from the stored past history,
while $\vec a_{y}^{+}$ is taken in the iterative scheme of sec.2.3  
from the trajectory of the previous iteration and in the 
direct integration scheme of sec.2.2 
as $\vec a_{y}=\dot{\vec v_{y}}/\dot y_{0}$, where $\dot{\vec v_{y}}$
is defined by backward differentiation formula). After these substitutions,
we have a closed system of differential-difference equations for
$\vec x(\tau),\vec y(\tau)$. Variables $x_{0}(\tau),y_{0}(\tau)$
do not participate in the equations, and can be determined 
by known solution of differential-difference equations using
recurrent formulae $x_{0}=y_{0}^{-}+|\vec y^{-}-\vec x|,\
y_{0}=x_{0}^{-}+|\vec x^{-}-\vec y|$.

\vspace{2mm}\noindent 3. 
The conserved quantities -- Noether's integrals of motion,
given by Eq.(2.9-10) in \cite{Schild}, after partial integration
can be written in the following form:

\begin{minipage}[c]{8cm}
\begin{eqnarray}
&&P_{\mu}=P_{\mu}^{(x)}+P_{\mu}^{(y)},\quad
P_{\mu}^{(x)}=m_{x}{\dot x_{\mu}\over\sqrt{\dot x^{2}}}
+{e_{x}\over2} A_{\mu}^{-}(x)\nn\\
&&+{e_{y}\over2} A_{\mu}^{+}(y^{-})
-\half(A^{-}(x)A^{+}(y^{-}) )(x-y^{-})_{\mu}\nn\\
&&-G_{\mu}^{(x)}(x)+G_{\mu}^{(y)}(y^{-}),
\quad P_{\mu}^{(y)}=(x \leftrightarrow y),\nn\\
&&G_{\mu}^{(x)}(\xi)=
{e_{x}\over2}
\int_{0}^{\xi}d\tau_{x}\dot x_{\nu}F_{\nu\mu}^{+}(x),
\quad G_{\mu}^{(y)}(\xi)=(x \leftrightarrow y).\nn
\end{eqnarray}
\end{minipage}\quad\quad\quad\begin{minipage}[c]{3cm}
\begin{figure}\label{noether}
~\epsfysize=3cm\epsffile{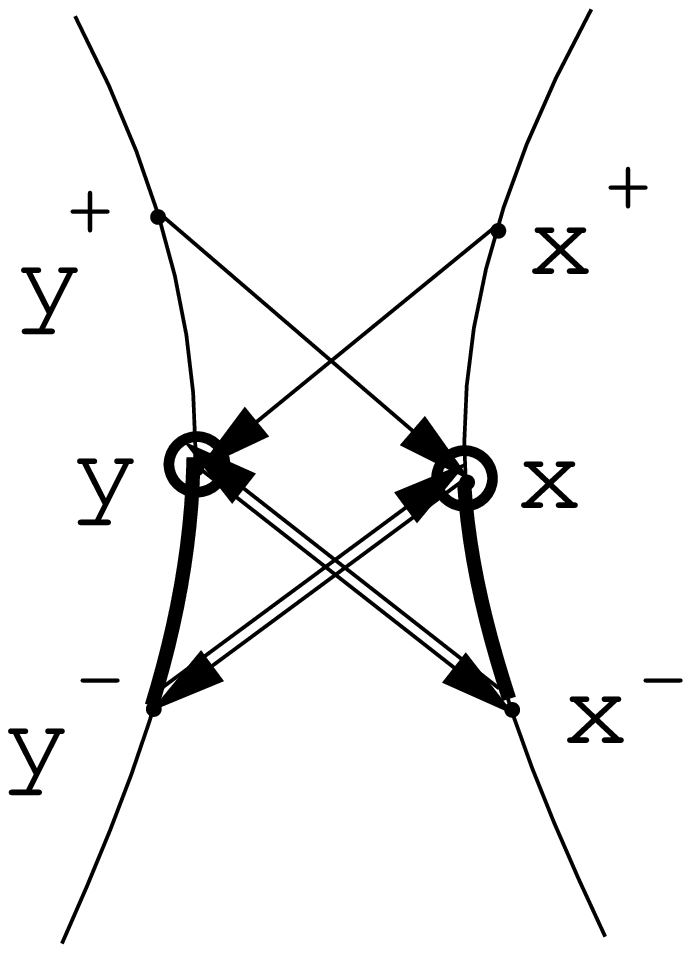}

\fignum Definition of Noether's integrals.
\end{figure}
\end{minipage}

\vspace{1mm}\noindent
Here $A_{\mu}(\xi)$ everywhere means 
vector-potential {\it felt} in point $(\xi)$. 
For example, $A_{\mu}^{\pm}(x)$ are the advanced and the retarded 
components of Lienard-Wierchert potential, created by particle $y$, 
felt in point $x$. The explicit formula for the potentials:
$A_{\mu}^{\pm}(x)=\pm e_{y}\dot y_{\mu}^{\pm}/
(r_{x}^{\pm}\dot y^{\pm})$. 
Analogously, $F_{\mu\nu}^{\pm}(x)$ is the field tensor,
created by particle $y$, felt in point $x$, given by formula
$F_{\mu\nu}^{\pm}(x)={\df A_{[\mu}^{\pm}(x)/\df x_{\nu]}}$,
where square brackets denote antisymmetrization of indices. The above defined
$G$-functions are equal to the integrated advanced Lorentz forces, 
acting on particles $x$ and $y$. Using the fact that a sum of
advanced and retarded Lorentz forces
after integration give a local expression (change of momentum),
$G$-terms can be rewritten to various equivalent forms,
{\em however} it is not posible to reduce them
completely to a local form (integrals cannot be eliminated). 
In our numerical scheme
the $G$-integrals are computed in parallel with 
the main integration processes, 
by the same high order scheme. The given expression for Noether's integrals
is graphically displayed by a diagram \fref{noether},
where the circles mean $m_{x,y}$-terms, solid lines are the segments
of integration in $G$-terms, and arrows the denote potentials $A^{\pm}$
(directed from source to destination). 

This representation has an important difference 
from the Noether's integrals of the non-relativistic case (Galilean):
The total momentum is composed of two contributions
$P_{\mu}^{(x)}$ and $P_{\mu}^{(y)}$, where $P_{\mu}^{(x)}$
depends on the position of the triple of points $(y^{-}xy^{+})$
and $P_{\mu}^{(y)}$ depends on the triple $(x^{-}yx^{+})$.
These triples {\it are independent}, as a result, 
both contributions are conserved separately:
$P_{\mu}^{(x)}=Const,\quad P_{\mu}^{(y)}=Const.$
It's easy to check that differentiation of $P_{\mu}^{(x,y)}$ 
gives separate equations of motion for particles $x$ and $y$. 

The expression for the angular momentum tensor
also consists of two separately conserved contributions
$L_{\mu\nu}=L_{\mu\nu}^{(x)}+L_{\mu\nu}^{(y)}$:

\begin{eqnarray}
&&L_{\mu\nu}^{(x)}=x_{[\mu}V_{\nu]}^{(x)}
+{e_{y}\over2}y_{[\mu}^{-} A_{\nu]}^{+}(y^{-})
-G_{\mu\nu}^{(x)}(x)+G_{\mu\nu}^{(y)}(y^{-}),
\quad L_{\mu\nu}^{(y)}=(x \leftrightarrow y),\nn\\
&&V_{\nu}^{(x)}=m_{x}{\dot x_{\nu}\over\sqrt{\dot x^{2}}}
+{e_{x}\over2} A^{-}_{\nu}(x)+\half(A^{-}(x)A^{+}(y^{-}) )y_{\nu}^{-},
\quad V_{\nu}^{(y)}=(x \leftrightarrow y),\nn\\
&&G_{\mu\nu}^{(x)}(\xi)={e_{x}\over2}
\int_{0}^{\xi}d\tau_{x}\dot x_{\rho}x_{[\mu}F_{\rho\nu]}^{+}(x),
\quad G_{\mu\nu}^{(y)}(\xi)=(x \leftrightarrow y).\nn
\end{eqnarray}

The above defined Noether's integrals are used to define the 
center-of-mass frame (CMF). 
Particularly, $L_{\mu\nu}$ defines the world line of CMF
origin by Eq.(2.13) in \cite{Schild}:
$c_{\mu}(\lambda)=L_{\mu\nu}P_{\nu}/P^{2}+\lambda P_{\mu}.$
In the coordinate system where the time axis is directed along this line
and the origin is located on this line with $c_{i}=0$,
three components of the angular momentum tensor form the angular
momentum vector: $\vec L=(L_{23},L_{31},L_{12})$
and the other three components vanish: $L_{0i}=0$. 
Check of this property, as well as conservation of 
Noether's integrals were used to control the precision 
of numerical analysis (see Appendix~C).

\paragraph*{Appendix B: Solving the equations of motion with respect 
to the most advanced variables} ~~~

\vspace{2mm}\noindent 1.
From equation (\ref{A1}) we see that the advanced contribution 
to the acceleration $\vec a_{x}$ is linear on the electric field 
$\vec E_{x}^{+}$:

$$a_{xi}^{(+)}=A_{ij}E_{xj}^{+},\ 
A_{ij}(y^{+};x,v_{x})={{e_{x}}\over{m_{x}}}(1-v_{x}^{2})^{1/2}
((1-nv_{x})\delta_{ij}+(n_{i}-v_{xi})v_{xj}),$$
where $\vec n=\vec r_{x}^{+}/|\vec r_{x}^{+}|,\ 
\vec r_{x}^{+}=\vec y^{+}-\vec x$. An analogous expression 
can be written for $\vec a_{x}^{(-)}$.
It is possible to prove that matrix $A$ is
non-degenerate, and that $\vec E_{x}^{+}$ can be unambiguously 
found as $\vec E_{x}^{+}=A^{-1}\vec a_{x}^{(+)}$.
Taking $\vec a_{x}^{(+)}=2\vec a_{x}-\vec a_{x}^{(-)}$, we finally express 
$\vec E_{x}^{+}$ as a function of the ``future'' variable $y^{+}$;
present variables $x,v_{x},a_{x}$ and past variables
$y^{-},v_{y}^{-},a_{y}^{-}$.

\noindent 2. 
According to (\ref{A1}), the electric field $\vec E_{x}^{+}$ 
depends on the position of the source $y^{+}$, its velocity $v_{y}^{+}$ 
and its acceleration $a_{y}^{+}$, and has the following cartesian components:
\begin{eqnarray}
&&E_{xi}^{+}={{-e_{y}}\over{r_{x0}^{+}(1-nv_{y}^{+})^{3}}}\left[
{{1-(v_{y}^{+})^{2}}\over{r_{x0}^{+}}}(n-v_{y}^{+})_{i}+\underbrace{\left(
(1-nv_{y}^{+})\delta_{ij}-(n-v_{y}^{+})_{i}n_{j}\right)}_{B_{ij}}a_{yj}^{+}
\right]\nn
\end{eqnarray}
Matrix $B$ is degenerate, i.e. $n_{i}B_{ij}=B_{ij}(n-v_{y}^{+})_{j}=0$. 
In order to solve this degenerate linear system 
with respect to $a_{y}^{+}$, we have to comply with the following:

a) condition of consistency (constraint found multiplying the system 
by $n_{i}$):
\begin{equation}
\underbrace{{{r_{x0}^{+~2}}\over{-e_{y}}}(E_{x}^{+}n)}_{C}
={{1-(v_{y}^{+})^{2}}\over{(1-nv_{y}^{+})^{2}}}
\label{constr}
\end{equation}

b) When this condition is satisfied, the solution can be written as
\begin{equation}
\vec a_{y}^{+}=-{{r_{x0}^{+}}\over{e_{y}}}(1-nv_{y}^{+})^{2}\vec E_{x}^{+}
+\lambda(\vec n-\vec v_{y}^{+})\nn
\end{equation}
where $\lambda$ is a so far arbitrary parameter. In the 2-dimensional  case, 
equation $(1-(v_{y}^{+})^{2})/(1-nv_{y}^{+})^{2}=C$ with $C>0$ defines 
an ellipses lying inside the circle $(v_{y}^{+})^{2}<1$; and with $C<0$ --
ellipses ($-1<C<0$), parabola ($C=-1$) and hyperbolas ($C<-1$), 
lying outside this circle.
The physical region corresponds to $C>0$, which means $|\vec v_{y}^{+}|<1$.
The consequence of the above is that it is impossible to solve uniquely for
the advanced accelerations $a_{y}^{+}$ from the equations of motion. Instead, 
we find a one-parameter degenerate solution plus one constraint:
equation (\ref{constr}) imposed on the velocity. In the following\cite{www} 
we choose a convenient set of variables to resolve the constraint 
and make the system available for explicit integration, 
then perform the integration using an adaptive delay equations 
integrator\cite{ChrisPaul}. In the result we find instability 
described in section 2.1. 

\begin{figure}\label{ell}
\begin{center}
\parbox{5cm}{\fignum Algebraic constraint.}
\quad\parbox{3cm}{
~\epsfysize=3cm\epsffile{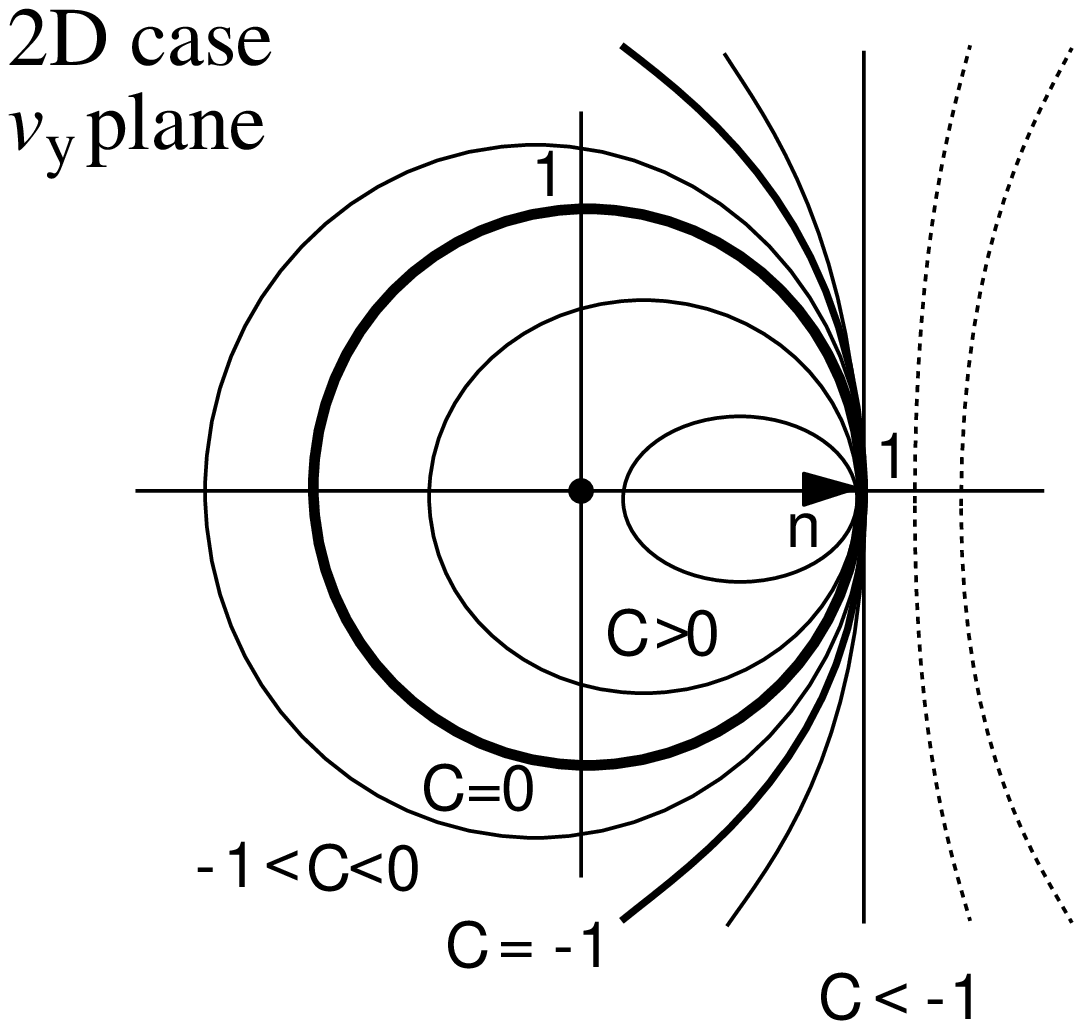}}
\end{center}
\end{figure}

\paragraph*{Appendix C: Details on the statement 
of the numerical experiments} ~~~

\vspace{2mm}\noindent 1. In the direct integration scheme
the initial segments of trajectories were set
to circular orbits \cite{Schild}.
This setting made in one half of one ladder step 
$\tau\in[0,0.5]$ is sufficient to define the further motion. 
However, using St\"urmer's method of integration, we need also to initialize
the finite differences $\Delta^{p}q_{k}$. Due to this reason we
are setting initial segments of trajectories to circular orbits
on longer intervals (up to 8 integration steps). 
After this setting the integrator recovers the correct circular motion.
To consider non-circular orbits, at a short time after the initial setting 
we apply to one particle a local external force $F_{ext}$. 
As a result of this action, the system acquires non-zero total momentum
and starts to move away from the origin. We then transform back to CMF and
observe the evolution there.

\vspace{3mm}
\begin{figure}\label{ini}
\begin{center}
\parbox{5cm}{\fignum Initial segment of trajectory:
a) in direct integration scheme; b) in iterative scheme.}
\quad\parbox{7cm}{
a)~\epsfysize=1.5cm\epsffile{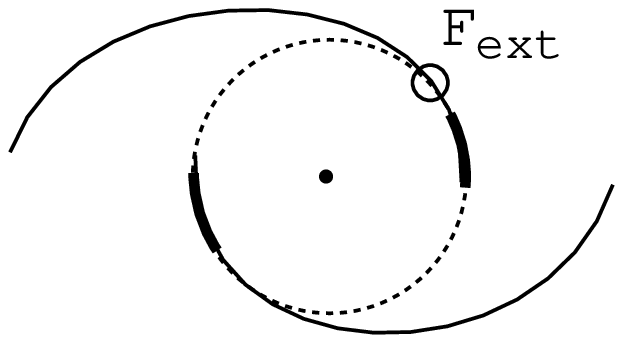}
~b)~\epsfysize=1.5cm\epsffile{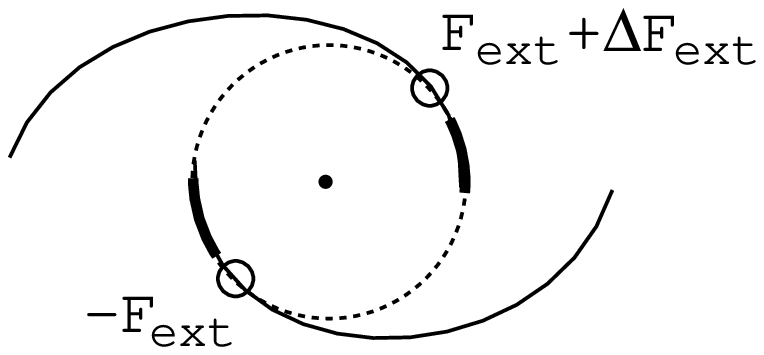}}
\end{center}
\end{figure}

\vspace{3mm}
Using St\"urmer's formulae of 8th order (\ref{Stur8})
and the corresponding backward differentiation formula (\ref{BDF}),
we have the acceleration, estimated with a precision of $o(h^{6})$.
The right hand side of the differential 
equations $f(t_{k},x_{k},...)$ are dependent on the computed acceleration, 
and it also has precision $o(h^{6})$.
As a result, in the 8th order St\"urmer's formula: 
$x_{k+1}=x_{k}+hf_{k}+...+o(h^{8})$
the last term ($\Delta^{7}q_{k-7}\sim h^{8}$) becomes ineffective,
actually the method has 7th order. However, the term $\Delta^{7}q_{k-7}$
should be in any case computed to find the acceleration
and can be preserved in this formula. 
Note also, that the acceleration terms in the Lorentz force
are suppressed by additional small factor $e^{2}/mr=r_{0}/r$
(ratio of classical radius to particles separation)
so that true precision of the method is between $o(h^{8})$ 
and $o((r_{0}/r)h^{7})$.

Using the grid of the light ladder parameter $\tau=n\in{\bf Z}$,
we still have convergent scheme, until the higher order differences 
$\Delta^{k}q_{n-k}$ are small. Actually, the role of parameter $h$ here
is played by retardation angle $\theta\sim v/c$. Near circular orbits
we have an estimation $r_{0}/r\sim(v/c)^{2}$, so that for $v/c\sim10^{-2}$ 
the application of 8th order scheme gives the error of order $10^{-16}$, 
on the level of machine double precision.

The integration process can be considerably accelerated by means
of {\it caching} of previously computed data. For this purpose
we store a short buffer of past data (8 integration steps), 
including all variables, whose evolution is tracked,
as well as the computed right hand sides of the equations of motion $q_{k}$,
their finite differences $\Delta^{p}q_{k}$ and some auxiliary variables
(such as $r_{x,y}^{\pm}$, which due to relations 
$r_{x}^{\pm}(\tau)=-r_{y}^{-}(\tau\pm0.5)$ enter in computation
several times). The content of the buffer is periodically written to a file 
to allow restarting of the program from previous states.
To transform the system to CMF we perform a Lorentz transformation 
of all variables in the buffer (some of these variables, e.g. accelerations, 
have non-linear transformation laws).

Control of the precision was performed by comparing two definitions of
acceleration (variable $\Delta\vec a$ in sec.2.2)  
and testing the conservation of Noether's integrals (Appendix~A). 
The relative difference $|\Delta\vec a|/|\vec a|$ was $<3\cdot10^{-10}$ 
in the region of free motion (in the region, where the external 
perturbation is applied, this value was about $0.07$, i.e. the method has 
worse precision in the region of perturbation, but further free motion
is integrated precisely). As already mentioned, Noether's integrals of 
Appendix~A separates in two conserved quantities 
$P_{\mu}=P_{\mu}^{(x)}+P_{\mu}^{(y)}$. Their conservation 
was satisfied in our numerical scheme (order $n=8$) 
with a precision of $|\Delta P_{0}^{(x,y)}|/P_{0}<10^{-11}$,
$|\Delta\vec P^{(x,y)}|/P_{0}<10^{-13}$. For total values in CMF we had
$|\vec P|/P_{0}<10^{-14}$. For the angular momentum tensor we had found 
that its components have a cumulative numerical error, 
and for these values in the CMF we have an estimation 
$|L_{0i}|/|L_{0i}^{(x)}|,|\Delta\vec L|/|\vec L|<10^{-14}*$num.of revolutions,
valid up to the maximal number of revolutions $=10^{6}$ we considered.

The direct integration scheme is applicable upto $v=0.006$ ($n=8$). 
For greater velocities the solutions exhibit oscillations
with time size comparable with the integration step,
and rapidly increasing amplitude. The appearance of these oscillations
is followed by violation of the conservation laws. 
This effect is caused by the increase of the light ladder step to a critical 
value, when it cannot be used anymore as a step of integration. 
The upper limit of $v$ creates on \fref{f0}c the right limit
of $E$ (high velocity almost circular orbits) and lower limit of $L$ 
(in this limit the trajectories are thin ellipses, and
the particles come close to each other, reaching the upper limit of $v$). 

In the considered range of velocities the precession of the orbits is so slow,
that the measurement of its angular frequency 
$\omega_{p}=$(angle of precession)/(time) requires special efforts.
For this purpose we find intersections of the orbits with the circles 
$r=Const$ and measure the difference of the angles $\Delta\alpha_i$ for 
consecutive intersection points. For better precision we interpolate the
solution between the integration points using the same high order
polynomial of the integration method and find the intersection
using dichotomy. Then we use large statistics of $\Delta\alpha_i$
to find the average $\Delta\alpha$, which is related to the frequences 
as $\Delta\alpha=2\pi(\omega_{p}/\omega_{0})$. Its 
mean-square error consists typically of about $10^{-13}$ radians,
leading to a relative error of $(\omega_{p}/\omega_{0})$
of about $10^{-9}$, sufficient for our purposes.

\vspace{2mm}
\begin{figure}\label{circ}
\begin{center}
\parbox{6cm}{\fignum Measurement of precession rate.}
\quad\parbox{4cm}{~\epsfysize=2cm\epsffile{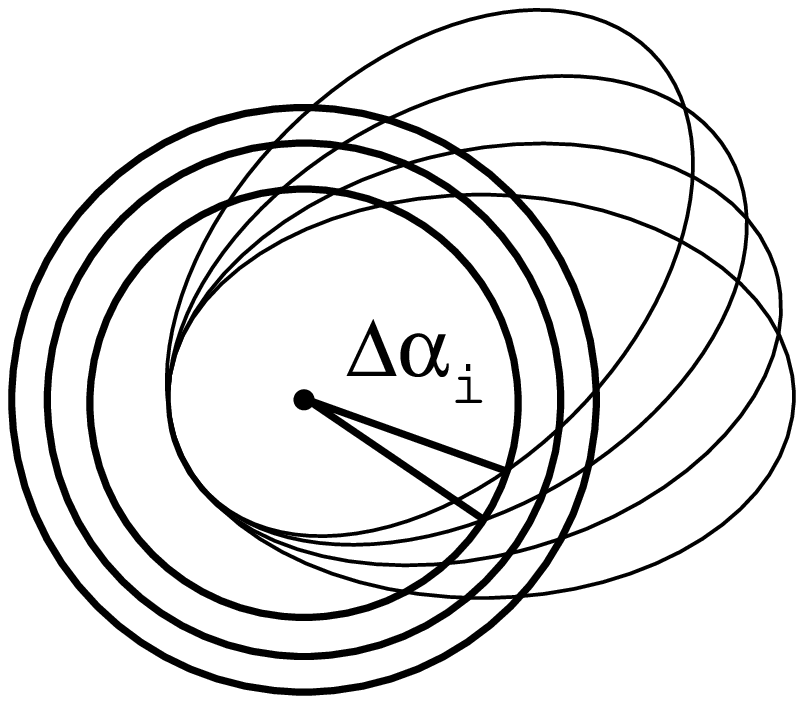}}
\end{center}
\end{figure}

\vspace{2mm}
Construction of the scatter plot \fref{f0}: The four main input 
parameters of the program are the radius of the initial circular orbit and 
the 3 components of the external perturbation force. These were varied in a 
certain 4-dimensional region and 10,000 points were generated in the run. 
In 25\% of cases the system was ionized (all these cases correspond to
$\sqrt{P^{2}}>2m$, i.e. when the external force gives the system 
enough energy for ionization). In 0.8\% of the cases the particles 
were too close to each other and moved too fast and as a result, 
the algorithm lost applicability and diverged. 
These cases were rejected at earlier stages of the integration 
and did not consume much computation time. The remaining cases 
are displayed on \fref{f0}. 

To prove that the scatter plot occupies a 2-dimensional surface
in the 4-dimensional space, we consider its sections by hypersurfaces 
$EL^{2}=Const$. The projection of these sections on the plane 
$(L,\Delta\omega_{p}/\omega_{0})$ is shown on the inner image of \fref{f0}b; 
projections to the planes $(E,\Delta\omega_{p}/\omega_{0})$ and $(\omega_{0},$
$\Delta\omega_{p}/\omega_{0})$ look similarly. As a result, we can conclude
that the sections are 1-dimensional curves in a 4-dimensional space,
which span a 2-dimensional surface with a change of the section parameter.
Two other approaches to measure the dimension of the scatter plot 
are described in \cite{www}: (1) computation of the Jacoby matrix for the 
mapping from the space of input parameters to the space of measured 
characteristics, and estimating its rank; (2) direct visualization of this 
4-dimensional object by projection to the 3-dimensional space with color 
encoding of the 4th coordinate.

\vspace{2mm}\noindent 2. In the iterative scheme the initial segments 
of the trajectories $\tau\in[0,0.5]$ were set to circular orbits.
To obtain non-circular orbits, we apply to both particles
the local external forces $-\vec F_{ext}$ and 
$\vec F_{ext}+\Delta\vec F_{ext}$. Such setting is convenient 
to control separately the orbits deviation from circular 
and asymmetry of initial conditions (the first one is controlled by
$\vec F_{ext}$, the second by $\Delta\vec F_{ext}$). 

\vspace{1mm}\noindent {\it Note:} 
There is still another type of the initial condition,
where the data on initial segment are directly set to non-circular orbits. 
In principle, these two types of initial conditions are equivalent:
the equations of motion are generally not satisfied on the initial segment,
where the trajectories {\it are set} to arbitrary shapes, 
and non-zero right hand side of the equations in this segment play 
the role of an effective external force. However, concrete implementation
of the non-circular initial condition creates certain problems in our scheme.
The data on the initial segment should necessarily be presented 
in light ladder parametrization, for which the analytical representation
is known only for the case of circular orbits. Moreover, the correct 
parametrization on the whole trajectory is recovered by further integration
process from the initial segment, and its $C^{n}$-smoothness is guaranteed
only if special conditions are satisfied (left and right derivatives 
upto $n$th order should coincide in a point $\tau=0.5$). For non-circular
initial orbits the satisfaction of this property is problematic,
while for circular orbits a global $C^{\infty}$-smooth parametrization 
can be easily constructed, and infinitely-differentiable external forces, 
used to control the shape of solution, preserve smoothness. 

For convergence of the method the total number of light ladder steps 
$Nls$ and the number of integration steps per light ladder step $Nps$ 
should be set to $Nps>10^{2},Nls<10$ at intermediate energies $E\sim1$ 
and to $Nps>10^{3},Nls<5$ at high energies $E\sim3$. Our stop-criterion 
of iteration was $\sum_{i}((\vec x_{n}(\tau_{i})-\vec x_{n-1}(\tau_{i}))^{2}
+(\vec y_{n}(\tau_{i})-\vec y_{n-1}(\tau_{i}))^{2})/(Nps*Nls)<\epsilon$,
where $(\vec x_{n},\vec y_{n})$ represent the current iteration,
$(\vec x_{n-1},\vec y_{n-1})$ -- previous iteration, the sum is taken
over all integration points, and $\epsilon=10^{-12}$.

In the described simplest form the method can be applied up to velocities 
$v=0.8\ (E\sim1)$, then a stabilization is needed to keep it convergent.
A stabilization procedure, proposed in \cite{Hans}, suggests to combine
the acceleration computed by the equations of motion 
with the acceleration stored from the previous iteration 
to $\vec a_{n}(t)=0.1\vec a_{eq}(t)+0.9\vec a_{n-1}(t)$,
and integrate it to the trajectories of the current iteration.
We have found that this type of stabilization, being applied
to our problem at high energies, makes the method indifferently stable: 
it has no divergence but cannot reach the solution either. 
We have used another approach, based on the following
predictor-corrector algorithm: We track the solution 
as a function of one control parameter $\mu$ (which, for example, 
can be equal to initial velocity of the particles). 
The solutions found for the $n$ previous values of $\mu$ should be stored 
($n$ old copies of coordinate, velocity and acceleration arrays
should be kept; while other variables, entering in the caching buffer, 
should not be copied). The dependence of the solution on $\mu$ 
at each integration point is approximated by a polynomial 
of $k$th order using a least square method (in our implementation $n=10,k=2$).
This polynomial is used as a starting point for the iterations 
at a new value of the control parameter $\mu+\Delta\mu$. 
If the solution has a smooth dependence on the control parameter,
this method produces a starting point placed much closer to the result
then other types of initial guess. This fact considerably improves 
the convergence. The step $\Delta\mu$ is chosen adaptively: 
decreased twice if convergence is lost and increased twice 
if the solution has been found, automatically keeping this value 
in the optimal region. This approach, earlier applied in \cite{fw_jmpc0}
for the solution of 1D WF, accelerates the recovery
of the solution along the $\mu$-axis by a factor of thousand, and keeps
the method stable up to the velocity $v=0.98$. 

Other parameters, influencing the shape of solution, such as the components 
of the external force, in this method should be set to the given 
smooth functions of $\mu$, which should change from zero to (for example) 
certain constant values. Starting the procedure for circular orbits 
at the velocity $v=0.8$, where the simplest scheme is convergent, 
we track one-dimensional families of solutions, corresponding 
to the given functions, and finally cover the whole space of 
input parameters in the region of convergence of the method. 
On \fref{xy1} the right limit in $E$, 
bottom limit in $\Delta L$ and upper limit in $\eta^{2}$ 
are the limits of convergence of the method respectively 
at high energy, strong perturbation force $F_{ext}$ and large
asymmetry $\Delta F_{ext}$. The upper limit in $\Delta L$, close to 
the circular orbit, is related with a problem of another kind: Such solutions
have very good convergence, but do not exhibit linear dependence 
in graphs $\eta^{2}(E)$, so that we cannot clearly separate 
symmetric and asymmetric phases by the described method.

\begin{figure}\label{f4}
\begin{center}
\parbox{5cm}{\fignum Two types of boundary effects 
in iterative scheme.}\quad\parbox{6cm}{
a)~\epsfysize=2cm\epsffile{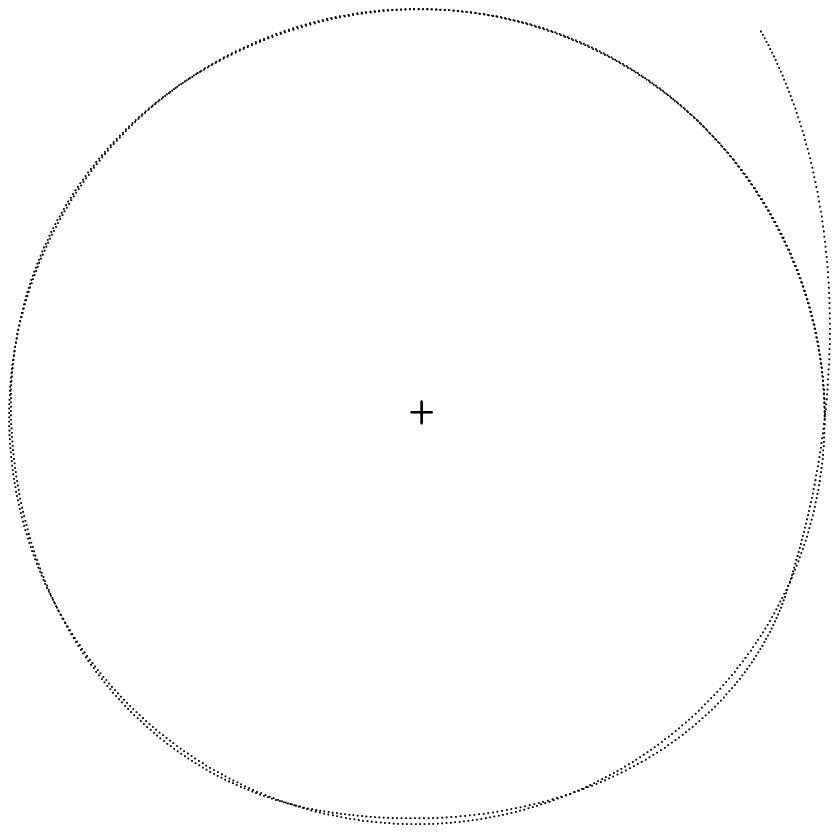}
\quad~b)~\epsfysize=2cm\epsffile{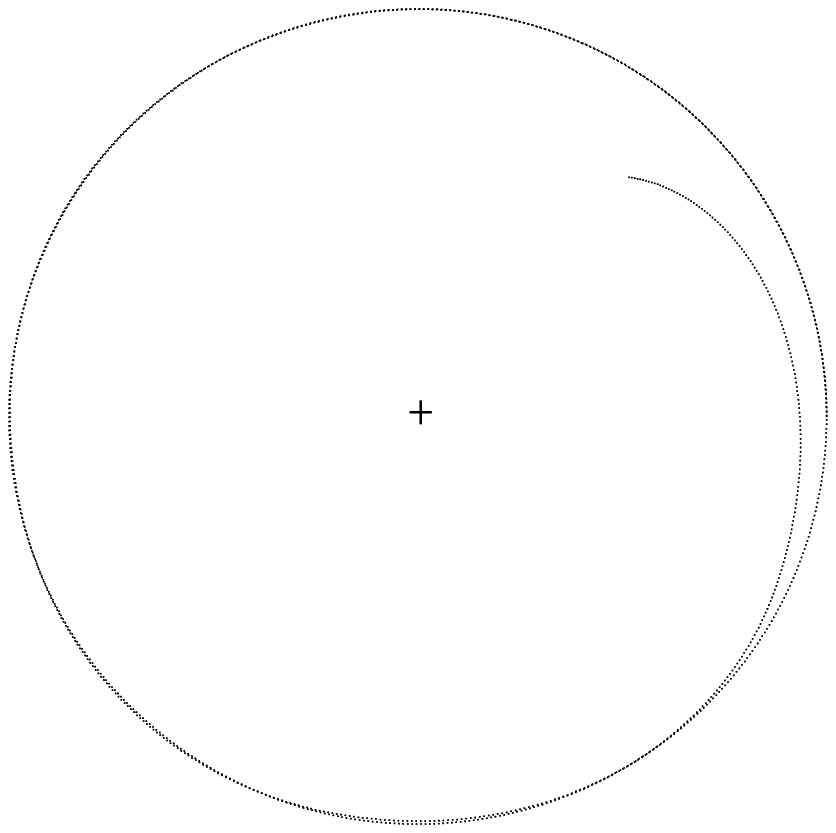}}
\end{center}
\end{figure}

An extra feature of the iterative scheme is the existence of artificial 
boundary effects, appearing near the end of the data arrays.
Because at the end of the integration interval the future evolution is not 
known, the advanced Lorentz force should be omitted in this region.
For the retarded Lorentz force we have used two possibilities: 
(a) single retarded Lorentz force or (b) double retarded Lorentz force 
is taken on the last step of the light ladder.
Absence of the advanced force and various settings for retarded force
lead to boundary effects, propagating to the inner regions of the integration 
interval. The amplitude of these effects is exponentially decreasing 
with the depth of penetration, so that the solution at several light 
ladder steps before the end of the integration is no more sensitive 
to boundary effects. It has also been found that boundary conditions 
of type (a) lead to stronger boundary effects, but support
better convergence at higher energies than condition (b).
Fig.\ref{f4} shows the solutions, corresponding to boundary conditions
of both types for value of velocity $v=0.86$. 

\vspace{2mm}\noindent The described modification 
of the Lorentz force leads to non-conservation of the integrals of motion 
in the last step of light ladder. At the beginning of the trajectory
in those regions, where the external forces are applied
(and where the initial data are set -- as we show above, 
this setting leads to the appearance of an effective external force) 
Noether's integrals are also not conserved. On the remaining parts 
of the trajectory the conservation laws are valid, namely these
parts are shown in the images \fref{xy}. The conservation laws
in these parts were satisfied with precision $|\vec P|/P_{0}<10^{-10},
|\Delta\vec L|/|\vec L|<10^{-8},|L_{0i}|/|\vec L|<10^{-7}$
at intermediate energies $E\sim1$, slowly increasing upto 
$|\vec P|/P_{0},|\Delta\vec L|/|\vec L|<10^{-6},
|L_{0i}|/|\vec L|<10^{-4}$ at high energies $E\sim3$.

\vspace{2mm}\noindent 3. Computational time:
most extensive computations are required
for the construction of the scatter plot \fref{f0} (24 hours) 
and 3D graph \fref{xy1} (120 hours). The computation was 
performed during one night, parallelly on 12 processors 
(300MHz MIPS R12000) of the GMD computer SGI/Onyx2.


\nonumsection{References}


\begin{thebibliography}{99}

\bibitem{fw_jmpc0} S. V. Klimenko, I. N. Nikitin, W. F. Urazmetov, 
{\it Int. J. Mod. Phys. C.} {\bf 10}, 905 (1999).

\bibitem{Fey-Whe}  J. A. Wheeler and R. P. Feynman, {\it Rev. of Mod. Physics%
}, {\bf 17}, 157 (1945) and {\it Rev. of Mod.} {\it Phys}. {\bf 21}, 425
(1949).

\bibitem{Schw-Tetr-Fokk}  K. Schwarzschild, Gottinger Nachrichten, 128,132
(1903), H. Tetrode, Zeits. f. Physisk {\bf 10}, 137 (1922) and A. D. Fokker,
\ Zeits. f. Physik {\bf 58}, 386 (1929).

\bibitem{Starusk} A. Staruskiewicz, {\it Ann. Physik} {\bf 25}, 362 (1970).

\bibitem{Anderson}  J. L. Anderson, {\em Principles of Relativity Physics },
Academic press, New York 1967, p.225.

\bibitem{Leiter}  D. Leiter, {\it Am. J. Phys}. {\bf 38}, 207 (1970).

\bibitem{Plass}  G. N. Plass, {\em PhD. Thesis}, Princeton University,1946
(UMI dissertation services).

\bibitem{Dirac}  P. A. M. Dirac, {\it Proc. Roy. Soc. London }{\bf 167}, 148
(1938).

\bibitem{Einstein}  A. Einstein and W. Ritz, {\it Phys. Z }{\bf 10}, 323
(1909).

\bibitem{Narlikar}  F. Hoyle and Jayant V Narlikar, {\em Cosmology and
Action at a Distance Electrodynamics, \ }(World Scientific, Singapore 1996),
see also F. Hoyle and J. V. Narlikar, {\it Rev. of Mod. Phys.} {\bf 67}, 113
(1995).

\bibitem{Schonberg}  M. Schonberg, {\it Phys. Rev.} {\bf 69}, 211 (1946).

\bibitem{Schild}  A. Schild, {\it Phys. Rev.} {\bf 131}, 2762 (1963) and
Science {\bf 138}, 994 (1962).

\bibitem{VonBaeyer}  C. M. Andersen and H. C. von Baeyer, {\it Phys. Rev. D} 
{\bf 5}, 2470 (1972).

\bibitem{Driver1}  R. D. Driver, {\it Phys. Rev.} D {\bf 19}, 1098 (1979), \
J. Hoag and R. D. Driver, {\it Nonlinear Analysis, Theory, Methods \&
Applications }{\bf 15}, 165 (1990).

\bibitem{Igor}  S. V. Klimenko, I. N. Nikitin and W. F. Urazmetov, {\it Il
Nuovo Cimento} {\bf 111}, 1281 (1998) and{\it \ Il Nuovo Cimento} {\bf 110},
771 (1995).

\bibitem{Hans}  C. M. Andersen and H. C. von Baeyer, {\it Phys. Rev. D} {\bf %
5}, 802 (1972). The sign of the acceleration terms in equation 2.3 of this 
paper is wrongly calculated and this might affect the stability results.

\bibitem{Harvard}  R. A. Moore, D. W. Qi and T. C. Scott, {\it Can. J. Phys.}
{\bf 70}, 772 (1992).

\bibitem{Elsgolts}  L. E. El'sgol'ts and S. B. Norkin, {\em Introduction to
the Theory and Application of Differential Equations with Deviating
Arguments, (}Academic Press, New York 1973).

\bibitem{Darwin} C. G. Darwin, {\it Phil. Mag.} V.39, p.537 (1920).



\bibitem{simple}  J. De Luca, Phys. Rev. E {\bf 62, } 2060 (2000).

\bibitem{Jackson}  J. D. Jackson, {\em Classical Electrodynamics}, second
edition, (John Wiley\&Sons, New York 1975), p.657.

\bibitem{ChrisPaul}C.A.H. Paul, {\em Numerical Analysis Report No. 283},
Manchester Centre for Computational Mathematics (1995) 
(http://www.ma.man.ac.uk/MCCM/MCCM.html).

\bibitem{www} homepage of the project: 
 http://viswiz.gmd.de/\~{}nikitin/wf/new/idea.html
\end{thebibliography}
\end{document}